\documentclass{aa}

\usepackage{ifthen}
\usepackage{graphicx,natbib,epsfig}

\bibpunct{(}{)}{;}{a}{}{,}

\def \ie{{\rm i.e.}}
\def \eg{{\rm e.g.}}

\def\cc{\ifmmode{\,{\rm cm}^{-3}}\else{$\,{\rm cm}^{-3}$}\fi}
\def\cq{\ifmmode{\,{\rm cm}^{-2}}\else{$\,{\rm cm}^{-2}$}\fi}
\def\mic{\ifmmode{\,\mu{\rm m}}\else{$\mu$m}\fi}
\def\eccs{\ifmmode{\,{\rm erg}\,{\rm cm}^{-3} {\rm s}^{-1}}\else{$\,{\rm
erg}\,{\rm cm}^{-3} {\rm s}^{-1}$}\fi}
\def\ecqs{\ifmmode{\,{\rm erg}\,{\rm cm}^{-2}\,{\rm s}^{-1}\,{\rm
sr}^{-1}}\else{$\,{\rm erg}\,{\rm cm}^{-2}\,{\rm s}^{-1}\,{\rm sr}^{-1}$}\fi}
\def\deg{\ifmmode{^{\circ}}\else{$^{\circ}$}\fi} 
\def\pc{\ifmmode{\,{\rm pc}}\else{$\,{\rm pc}$}\fi} 
\def\kms{\ifmmode{\,{\rm km}\,{\rm s}^{-1}}\else{km s$^{-1}$}\fi} 
\def\kmspc{\ifmmode{\,{\rm km}\,{\rm s}^{-1}\,{\rm pc}^{-1}}\else{km
s$^{-1}$ pc$^{-1}$}\fi} 
\def\MJysr{\ifmmode{\,{\rm MJy\,sr}^{-1}}\else{$\,{\rm MJy\,sr}^{-1}$}\fi} 
\def\Kkms{\ifmmode{\,{\rm K\,km\,s}^{-1}}\else{$\,{\rm K\,km\,s}^{-1}$}\fi} 
\def\twCO{\ifmmode{\rm ^{12}CO}\else{$\rm^{12}CO$}\fi} 
\def\thCO{\ifmmode{\rm ^{13}CO}\else{$\rm^{13}CO$}\fi} 
\def\CeiO{\ifmmode{\rm C^{18}O}\else{$\rm C^{18}O$}\fi} 
\def \Cp{\ifmmode{\rm C^+}\else{$\rm C^+$}\fi} 
\def \CHp{\ifmmode{\rm CH^+}\else{$\rm CH^+$}\fi}
\def \thCHp{\ifmmode{\rm ^{13}CH^+}\else{$\rm ^{13}CH^+$}\fi}
\def \CHtp{\ifmmode{\rm CH_2^+}\else{$\rm CH_2^+$}\fi} 
\def\CHthp{\ifmmode{\rm CH_3^+}\else{$\rm CH_3^+$}\fi} 
\def \HCOp{\ifmmode{\rm HCO^+}\else{$\rm HCO^+$}\fi} 
\def \HtOp{\ifmmode{\rm H_3O^+}\else{$\rm H_3O^+$}\fi} 
\def \HCfiN{\ifmmode{\rm HC_5N}\else{$\rm HC_5N$}\fi} 
\def\wat{\ifmmode{\rm H_2O}\else{$\rm H_2O$}\fi} 
\def \oxy{\ifmmode{\rm O_2}\else{$\rm O_2$}\fi} 
\def \HH{\ifmmode{\rm H_2}\else{$\rm H_2$}\fi}
\def \Jone{\ifmmode{\rm {(J=1--0)}}\else{{(J=1--0)}}\fi} 
\def\Jtwo{\ifmmode{\rm {(J=2--1)}}\else{{(J=2--1)}}\fi} 
\def\Jfo{\ifmmode{\rm {J=4--3}}\else{{J=4--3}}\fi} 
\def \Jon{\ifmmode{\rm {J=1--0}}\else{{J=1--0}}\fi} 
\def \Jtw{\ifmmode{\rm {J=2--1}}\else{{J=2--1}}\fi} 
\def \Jth{\ifmmode{\rm {J=3--2}}\else{{J=3--2}}\fi} 
\def\Jfi{\ifmmode{\rm {J=4--3}}\else{{J=4--3}}\fi} 
\def \Ta{\ifmmode{\rm T_A}\else{$\rm T_A$}\fi} 
\def \Tas{\ifmmode{\rm T_A^*}\else{$\rm T_A^*$}\fi} 
\def \Tmb{\ifmmode{\rm T_{mb}}\else{$\rm T_{mb}$}\fi} 
\def \Tr{\ifmmode{\rm T_r}\else{$\rm T_r$}\fi} 
\def \Trs{\ifmmode{\rm T_r^*}\else{$\rm T_r^*$}\fi}

\begin{document}

\title{Intermittency of interstellar turbulence: extreme velocity-shears and CO  
  emission on milliparsec scale.\thanks{Based on
    observations obtained with the IRAM Plateau de Bure interferometer
    and 30~m telescope. IRAM is supported by INSU/CNRS (France), MPG
    (Germany), and IGN (Spain).}}

\author{E. Falgarone \inst{1}, J. Pety \inst{1,2} \and P. Hily-Blant
  \inst{3}}

\institute{
LERMA/LRA, CNRS, UMR 8112, Ecole Normale Sup\'erieure \& Observatoire de Paris, 
24 rue Lhomond, F-75005 Paris, France
\email{falgarone@lra.ens.fr}
\and Institut de Radio Astronomie  Millim\'etrique, 300 rue de la Piscine, F-38406 Saint Martin
  d'H\`eres, France \\ \email{pety@iram.fr}
\and LAOG, CNRS UMR 5571, Universit\'e Joseph Fourier, BP53, F-38041 Grenoble, France\\
\email{pierre.hilyblant@obs.ujf-grenoble.fr} }

\date{Received 14 September 2008 / Accepted 29 September 2009}

\abstract
{}
{The condensation of diffuse gas into molecular clouds and dense cores
occurs at a rate driven largely by turbulent  
  dissipation. This process still has
  to be caught in action and characterized.}
{We observed a mosaic of 13 fields with the IRAM-PdB interferometer (PdBI) to
  search for small-scale structure in the \twCO(1-0) line emission of
  the turbulent and translucent environment of a low-mass dense core in the Polaris
  Flare.  The large size of the mosaic (1'$\times$2') compared to the
  resolution (4'') is unprecedented in the study of the
  small-scale structure of diffuse molecular gas. }
{The interferometer data uncover eight weak and
  elongated structures with thicknesses as small as $\approx$ 3 mpc (600
  AU) and lengths up to 70 mpc, close
  to the size of the mosaic.  These are not filaments because 
once merged 
  with short-spacings data, the PdBI-structures appear to be the
  sharp edges, in space and velocity-space, of larger-scale structures.
Six out of eight form quasi-parallel pairs  
 at different velocities and different position angles.
This cannot be the result of chance alignment. The velocity-shears 
estimated for the three pairs include the highest values 
ever measured in regions that do not form stars (up to 780 \kmspc).
The CO column density of the PdBI-structures 
 is in the range $N({\rm CO)}=10^{14}$ to $10^{15}$\cq\ and their
 \HH\ density, estimated in several ways, does not exceed a few
  10$^3$ \cc. 
Because the larger scale structures
have sharp edges (with little or no overlap for those that are 
pairs), they have to be thin layers 
of CO emission. We call them SEE(D)S for Sharp-Edged  Extended (Double) Structures.
These edges mark a transition, on the milliparsec scale, between 
a CO-rich  component and a gas undetected in the \twCO(1-0) line because of its low CO 
abundance, presumably the cold 
neutral medium.  }
{We propose that these SEE(D)S are 
 the first directly-detected manifestations of the intermittency of interstellar turbulence.
The large velocity-shears reveal an intense straining field, responsible for a local  
dissipation rate several orders of magnitude above average, possibly at the 
origin of the thin CO layers. }

\authorrunning{Falgarone et al.}
\offprints{E.~Falgarone}
\titlerunning{Extreme velocity-shears and CO on milliparsec scale}

\maketitle

\section{Introduction}

\begin{figure}
  \centering
  \includegraphics[width=\hsize{}]{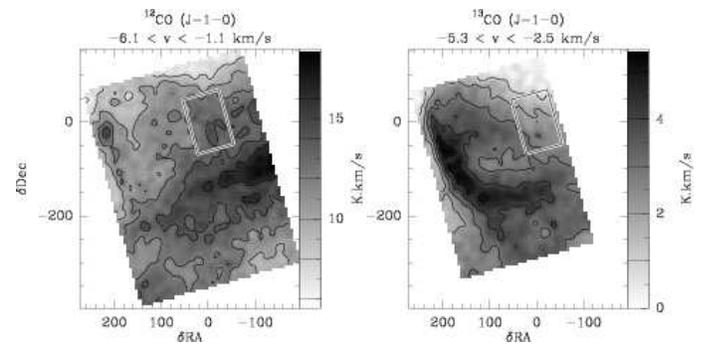}
  \caption{The location of the 13-field mosaic observed at the Plateau
    de Bure interferometer (centered at RA=01:55:12.26 and
    Dec=87:41:56.30) is shown as the box on top of the integrated
    emission of the \twCO{} and \thCO{} \Jone{} maps obtained at the
    IRAM-30m. This is a place of low, almost featureless, CO line
    brightness.  The arc-like structure visible in \thCO\ traces the outer layers of the low-mass 
dense core. Contour levels are shown in the wedges.}
  \label{fig:location}
\end{figure}

Turbulence in the interstellar medium (ISM) remains a puzzle 
in spite of dedicated efforts
on observational and numerical grounds.  
This is because it is compressible, magnetized, and
multi-phase, but also because of the huge range of scales
separating those of injection and dissipation of energy.  
Moreover, because turbulence and magnetic fields are the main
support of molecular clouds against their self-gravity, turbulent
dissipation is a key process among all those 
eventually leading to star formation \citep[see the reviews of][]
{elmescalo04,scaloelme04}.

In molecular clouds, turbulence is observed to be highly supersonic
with respect to the cold gas.  It is thus anticipated to dissipate in
shocks in a cloud-crossing time (\ie\ $\approx$ a few 10~Myr for
giant molecular clouds of 100~pc with internal velocity dispersion of
a few \kms). Magnetic fields do not
significantly slow the dissipation down \citep{mmml98}.  Actually, this is the basis of
the turbulent models of star formation \citep{maclow04} -- one of the two current
scenarii of low-mass star formation -- in which self-gravitating
entities form in the shock-compressed layers of supersonic turbulence.

However, while it is unquestionable that the ISM is regularly swept by large-scale 
shock-waves triggered by supernovae explosions that partly feed the 
interstellar turbulent cascade  \citep{joung06,avillez07}, 
the smallest scales, barely subparsec 
in these simulations, are still orders of magnitude above the smallest observed 
structures and are unlikely to provide a proper description of the actual 
dissipation processes. 
Whether turbulent dissipation occurs primarily in compressive
(curl-free) or in solenoidal (divergence-free) modes in  the interstellar medium 
has therefore to be considered as an open issue. 

An ideal target to study turbulent dissipation is the diffuse
molecular gas because it is the component in which dense cores form,
with less turbulent energy density than their environment.  
The word ``diffuse'' here comprises
all material in the neutral ISM at large that is not in dense cores
\ie\ whose total hydrogen column density
is less than a few $10^{21}$ \cq.  This includes the mixture of cold
and warm neutral medium (CNM and WNM), the edges of molecular cloud
complexes (also called translucent gas), and the high latitude clouds.
Diffuse gas builds up a major mass fraction of the ISM.  Actually, on the
  30~pc scale, \cite{goldsmith08} find that half the mass of the
  Taurus-Auriga-Perseus complex lies in regions having \HH\ column
  density below $2.1 \times 10^{21}$ \cq.

Turbulent dissipation may also provide clues to the  
``outstanding mysteries'' raised by observations of diffuse molecular gas 
\citep[see the review of][]{snow06}: the ubiquitous small scale
structure, down to AU-scales \citep{heiles07}, the remarkable
molecular richness found in this hostile medium, weakly shielded from
UV radiation \citep[e.g.][]{lilu98,gredel02},  the bright emission 
in the \HH\  pure rotational lines exceeding the predictions of  
photon-dominated region  (PDR)  models \citep{falgarone05,lacour05},
the \twCO\ small-scale structures with a broad
range of temperatures, \HH\ densities and linewidths
 that preclude a single interpretation in terms of cold
dense clumps \citep{ingalls2000, ingalls2007, heithausen04, heithausen06,
  sakamoto03}.

 The present paper extends the investigation of 
     of turbulence down to the mpc-scale in the translucent environment of a low-mass dense core of
  the Polaris Flare. Over the years, this investigation has progressed
  along three complementary directions: \\
  {\it (i)} A two-point statistical analysis of the velocity field
    traced by the \twCO\ line emission, and conducted on maps of increasing
    size. Using numerical simulations of mildly compressible turbulence,
    \cite{lis96} and~\cite{pety03} first proposed that the non-Gaussian
    probability distribution functions ({\it pdf}s) of line centroid
    velocity increments (CVI) be the signatures of the space-time
    intermittency of turbulence \footnote{ Intermittency here refers to the
      empirical property of high Reynolds number turbulence to present  an excess 
        of rare events compared to
       Gaussian statistics, this excess being increasingly large as
      velocity fluctuations at smaller and smaller scales are considered
      \citep[see the review of][]{anselmet01}. Although the origin of
      intermittency is still an open issue \citep[but
      see][]{mordant02,chevillard05,arneodo08}, it is quantitatively
      characterized by the anomalous scaling of the high-order structure
      functions of the velocity and the shape of non-Gaussian {\it pdf}s of
      quantities involving velocity derivatives \citep[e.g.][]{frisch95}.}
     because the extrema of CVI (E-CVI) trace
    extrema of the line-of-sight average of the modulus of the {\it pos}
    vorticity.  Statistical analysis conducted on parsec-scale maps in two
    nearby molecular clouds have revealed that these extrema form
      parsec-scale coherent structures~\citep[resp. Paper~III,
      HF09]{hily08,hf09}. \\
 {\it (ii)} A detailed analysis (density, temperature, molecular
    abundances) of these coherent structures, based on their molecular line emission.
     The gas there is more optically thin in the \twCO\ 
    lines, warmer and more dilute than the bulk of the gas \citep[hereafter Paper
    II]{hily07}, and  large \HCOp\ abundances, unexpected in an
    environment weakly shielded from UV radiation, have been detected there
    \citep[Paper~I]{falgarone06}.\\
   {\it (iii)} Chemical models of non-equilibrium warm chemistry
    triggered by bursts of turbulent dissipation~\citep{joulain98}.
  The most recent progresses along those lines include the chemical models
  of turbulent dissipation regions (TDRs) by \cite{godard09} and their successful
  comparison to several data sets, among which new submillimeter detections 
of \thCHp(1-0) (Falgarone et al. in preparation). 
 
 The \twCO\Jtwo\ 
  observations of the Polaris Flare with unprecedented angular resolution
  and dynamic range are the first to evidence the association between
  extrema of CVI and observed velocity-shears\footnote{We use
    velocity-shear rather than velocity-gradient because the observations
    provide cross-derivatives of the velocity field, \ie\ the displacement
    measured in the plane-of-the-sky ({\it pos}) is perpendicular to 
    the line-of-sight velocity} (HF09). No shock signature (density and/or
  temperature enhancement, SiO detection) has been found in the coherent
  structure of E-CVI identified in the Polaris Flare (Hily-Blant and
  Falgarone, in preparation).
All the above suggest (but does not prove yet) that the coherent
structures carrying the statistical properties of intermittency are
regions of intense velocity-shears where dissipation of turbulence is
concentrated.   

The \twCO(1-0) observations reported in this paper have been performed
in a field located on one branch of the Polaris Flare E-CVI structure, in the 
  translucent and featureless environment of a dense core
(Fig.~\ref{fig:location}).  The outline of the paper is the
following: the observations and data reduction are described in
Section 2. The observational results are given in Section 3. The
characterization of the emitting gas is made in Section 4 and we
discuss, in Section 5, the possible origin and nature of the CO
structures that we have discovered. Section 6 puts our results in the
broad perspective provided by other data sets and Section 7 compares
them to  chemical model predictions and numerical simulations of turbulence. 
The conclusions are given in Section 8.

\section{IRAM Plateau de Bure Interferometer observations}

We used the IRAM Plateau de Bure Interferometer (PdBI) to image, at
high angular resolution and in the \twCO{} \Jone{} line, a region of
$\sim 1'\times 2'$ in the translucent environment of a dense core in
the Polaris Flare \citep{heithausen99, heithausenBB02}.  The
location of the target field is shown in Fig.~\ref{fig:location} as a
rectangle on larger scale, single-dish maps of integrated
\twCO\Jone\ and \thCO\Jone\ emission from \cite[][hereafter F98]{falgarone98kp}. The average column
density in this region ($\sim 10^{21}$ \cq) is about 100 times smaller
than in the central parts of the dense core ($\sim 10^{23}$ \cq), 3
arcmin westwards.  The average integrated \thCO\ intensity over the
mosaic area is weak $W(\thCO)=2$ \Kkms.

\begin{table*}
  \caption{Observation parameters. The projection center of all the
    data displayed in this paper is: $\alpha_{2000} =
    01^h55^m12.26^s$, $\delta_{2000} = 87\deg41'56.30''$.}
  \begin{center}
    {\tiny
      \begin{tabular}{cccccccccc}
        \hline \hline
        Molecule & Transition & Frequency  & Instrument & Config. & Beam   & PA     & Vel. Resol. & Int. Time & Noise  \\
                 &            & GHz        &            &         & arcsec & $\deg$ & \kms{}      & hours     & K      \\
        \hline
        \twCO{} & \Jone{} & 115.271195 & PdBI & C\&D & $ 4.4 \times 4.2$ & 80 & 0.1 & 65.2/180$^{a}$ & 0.23$^{b}$ \\
                &         & 115.271195 &  30m &  --- &              21.3 &  0 & 0.1 &  ---/---       & 0.40       \\
        \thCO{} & \Jone{} & 110.201354 &  30m &  --- &              22.3 &  0 & 0.1 &  ---/---       & 0.19       \\
        \hline
      \end{tabular}}
  \end{center}
  \begin{list}{}{}
  \item $^{a}$ Two values are given for the integration time: the 5
    antennae array equivalent on-source time and the telescope time.
  \item $^{b}$ The noise value quoted here is the noise at the mosaic
    phase center
  \end{list}
  \label{tab:obs}
\end{table*}

\subsection{Observations}

The observations dedicated to this project were carried out in 1998
and 1999 with the IRAM interferometer at Plateau de Bure in the C and
D configurations (baseline lengths from 24m to 161m). One correlator
band of 10~MHz was centered on the \twCO{}~\Jone{} frequency to cover
a $\sim23\kms$ bandwidth with a channel spacing of 39~kHz, \ie{}
$\sim0.1\kms$. Four additional correlator bands of 160~MHz were used
to measure the 2.6~mm continuum over the 500 MHz instantaneous
IF-bandwidth then available.

We observed a 13-field mosaic centered on $\alpha_{2000} =
01^h55^m12.26^s$, $\delta_{2000} = 87\deg41'56.30''$. The field
positions followed a compact hexagonal pattern to ensure Nyquist
sampling in all directions and an almost uniform noise over a large
fraction of the mosaic area (see Fig.~\ref{fig:noise} of Appendix
A). The imaged field-of-view is about a rectangle of dimension of
$85'' \times 130''$ oriented at a position-angle of $15\deg$ (because
the (RA,Dec) PdBI field was selected in maps made in (l,b) coordinates).

Polaris being a circumpolar source, this project was a good
time-filler. It was thus observed at 22 different occasions, giving a
total of about 180~hours of \emph{telescope} time with most often 3 or
4 antennas and rarely 5 antennas. Taking into account the time for
calibration and data filtering this translates into \emph{on--source}
integration time of useful data of 65.2~hours for a full 5-antenna
array. The typical 2.6~mm resolution of these data is $4.3''$.  The
data used to produce the missing short-spacings are those of the IRAM
key-program, fully described in F98 (see also Table 1).

\subsection{Data reduction}

The data processing was done with the GILDASf\footnote{See
  \texttt{http://www.iram.fr/IRAMFR/GILDAS} for more information about the
  GILDAS{} softwares.} software suite~\citep{pety05}. Standard calibration
methods implemented in the GILDAS/CLIC program were applied using
close bright quasars as calibrators. The calibrated $uv$ tables were
processed through an Hanning filter which spectrally smoothed the data (to
increase the intensity signal-to-noise ratio) while keeping the same
velocity/frequency channel spacing.

All other processing took place into the GILDAS/MAPPING software.
Following~\citet{gueth96}, the single-dish map from the IRAM-30m key
program were used to create the short-spacing visibilities not sampled
at the Plateau de Bure. These were then merged with the
interferometric observations. Two different sets of $uv$ tables (\ie{}
with and without short-spacings) were then imaged using the same
method.  Each mosaic field was imaged and a dirty mosaic was built
combining those fields in the following optimal way in terms of
signal--to--noise ratio~\citep{gueth01}
\begin{displaymath}
\displaystyle J(\alpha,\delta) = \sum\nolimits_i
\frac{B_i(\alpha,\delta)}{\sigma_i^2}\,F_i(\alpha,\delta) \left/
\displaystyle \sum\nolimits_i \frac{B_i(\alpha,\delta)^2}{\sigma_i^2}.
\right.
\end{displaymath}
In this equation, $J(\alpha,\delta)$ is the brightness distribution in
the dirty mosaic image, $B_i$ are the response functions of the $i$
primary antenna beams, $F_i$ are the brightness distributions of the
individual dirty maps and $\sigma_i$ are the corresponding noise
values. As may be seen in this expression, the dirty intensity
distribution is corrected for primary beam attenuation, which makes
the noise level spatially heterogeneous.  In particular, noise
strongly increases near the edges of the field of view.  To limit this
effect, both the primary beams used in the above formula and the
resulting dirty mosaics are truncated. The standard level of
truncation is set at 20\% of the maximum in MAPPING.

Deconvolution methods were different for both data sets (\ie{} with
and without short-spacings). The dirty image of the PdBI-only data was
deconvolved using the standard Clark CLEAN algorithm. One spatial
support per channel map was defined by selecting positive regions on
the first clean image which was obtained without any constraint. This
geometrical constraint was then used in a second deconvolution. While
it can bias the result, this two-step process is needed when
deconvolving interferometric observations of extended sources without
short-spacings. Indeed, the lack of short-spacings implies (among
other things) a zero valued integral of the dirty beam and dirty
image, which in turn perturbs the CLEAN convergence when the source is
extended because the algorithm searches as much positive as negative
CLEAN components. The only way around is to guide the deconvolution by
the definition of a support where the signal is detected. On the other
hand, the deconvolution of the combined short-spacings and
interferometric $uv$ visibilities can process blindly (\ie{} without
the possible bias of defining a support where to search for CLEAN
components). This is what has been done and the good correlation of
the structures seen in the deconvolved images of the data with and
without short-spacings (see Fig.~4 and 5) gives us confidence in our
deconvolution of the PdBI-only data.

The two resulting data cubes (with and without short-spacings) were then
scaled from Jy/beam to \Tmb{} temperature scale using the synthesized beam
size (see Table~\ref{tab:obs}). Final noise rms measured at the centered of
the mosaic is about 0.23~K in both data cubes.

\section{Observational results}

\begin{figure}
  \centering
  \includegraphics[height=\hsize{},angle=270]{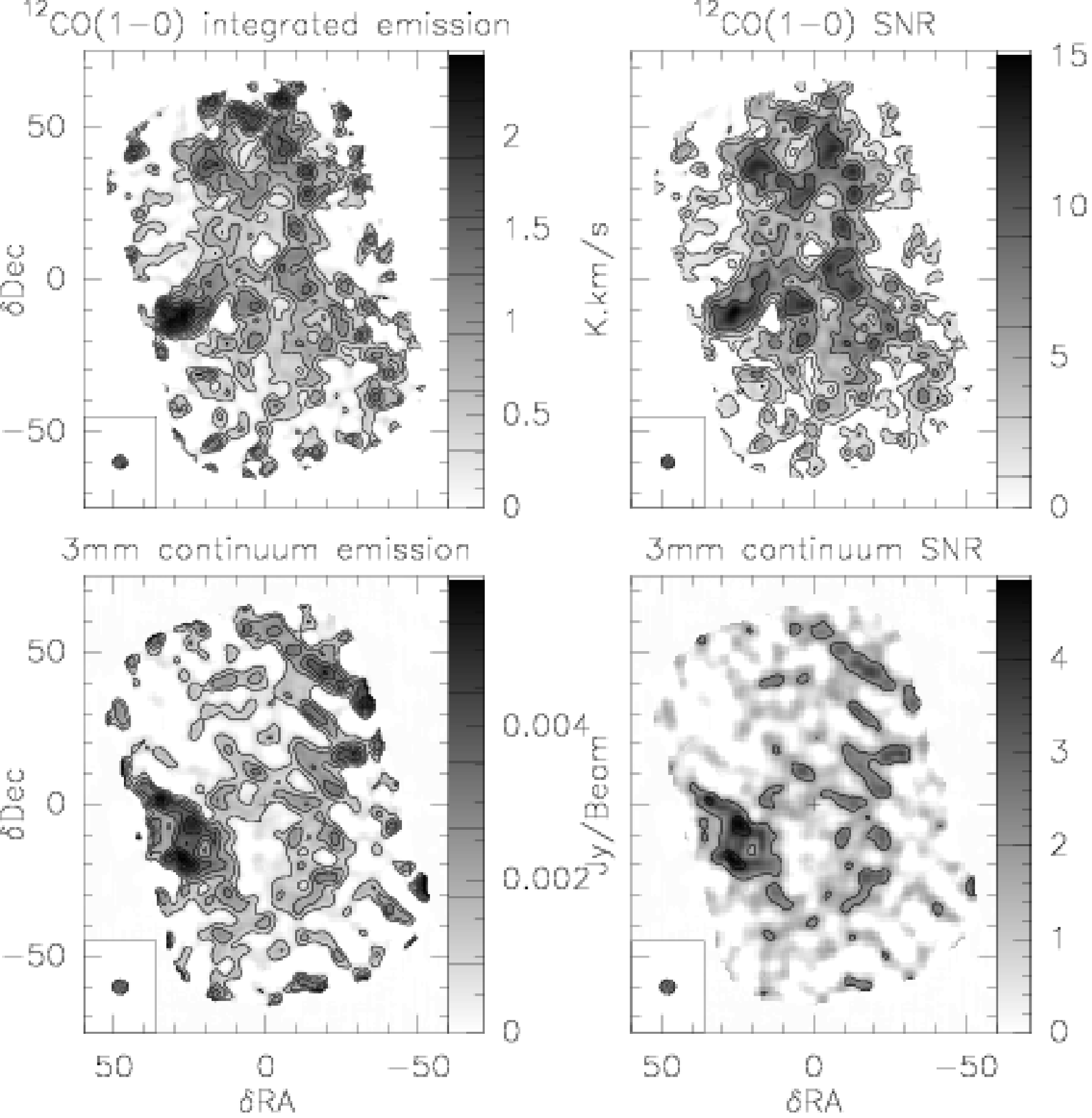}
  \caption{Map of integrated emission of the PdBI data (top left),
and signal-to-noise ratio (top right) for the
 \twCO{} \Jone{} line. Same for the 3mm continuum emission 
(bottom panels). The synthesized beam is shown in the bottom left inserts. } 
  \label{fig:integrated}
\end{figure}

\begin{figure*}
  \centering
 \includegraphics[height=\hsize{},angle=270]{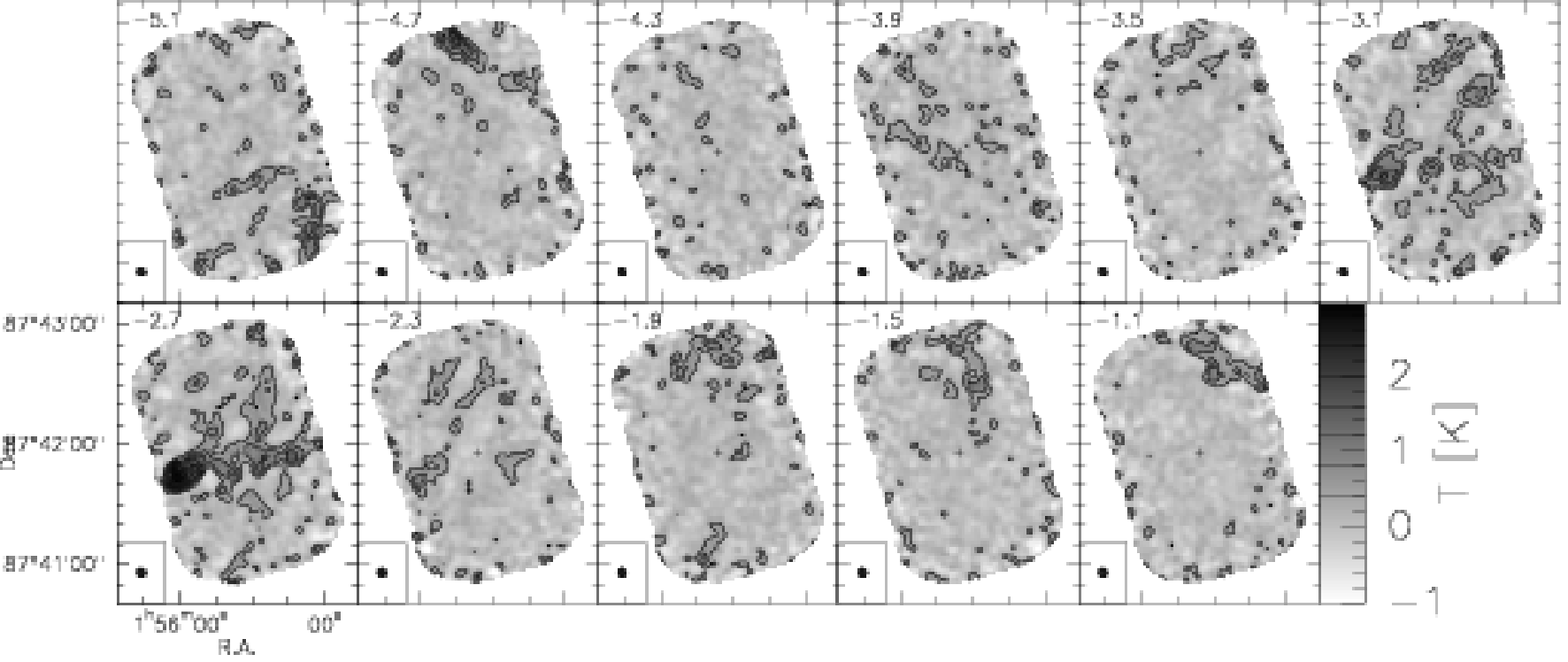}
  \includegraphics[height=\hsize{},angle=270]{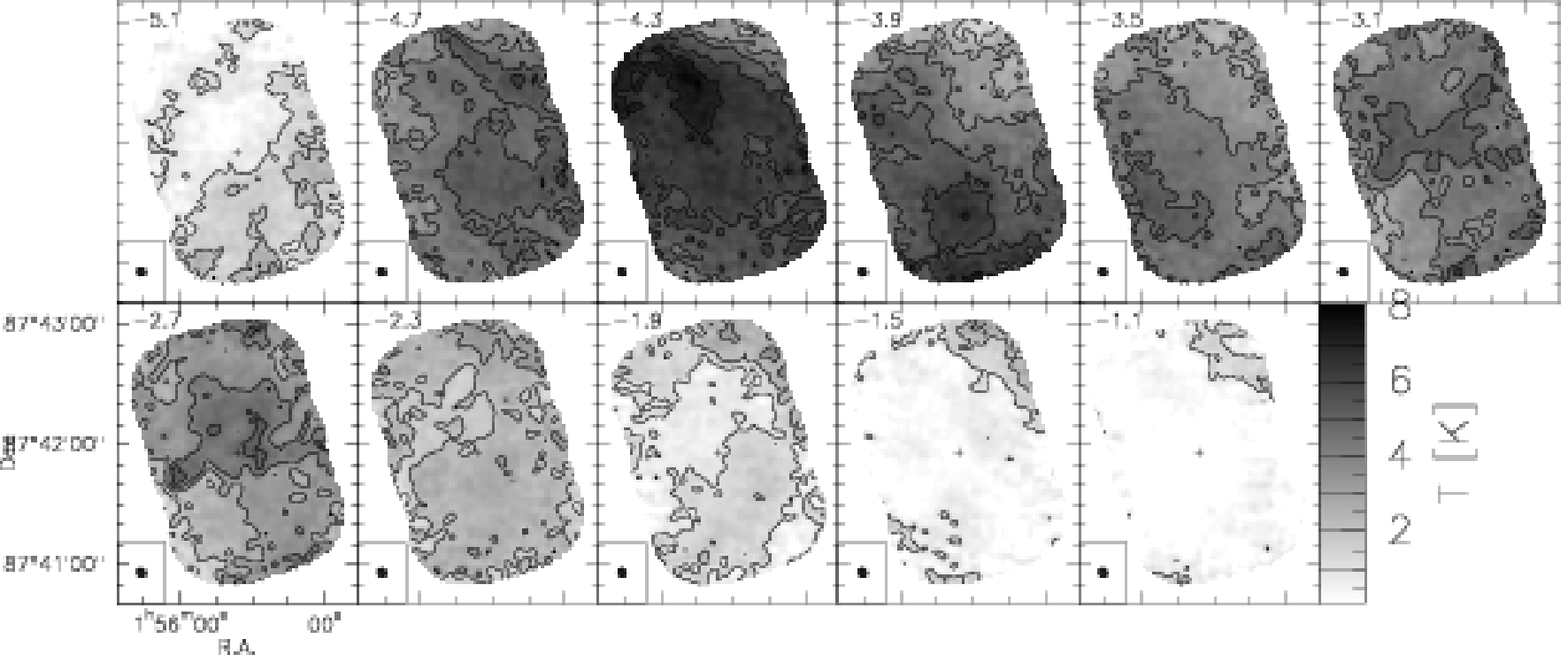}
  \includegraphics[height=\hsize{},angle=270]{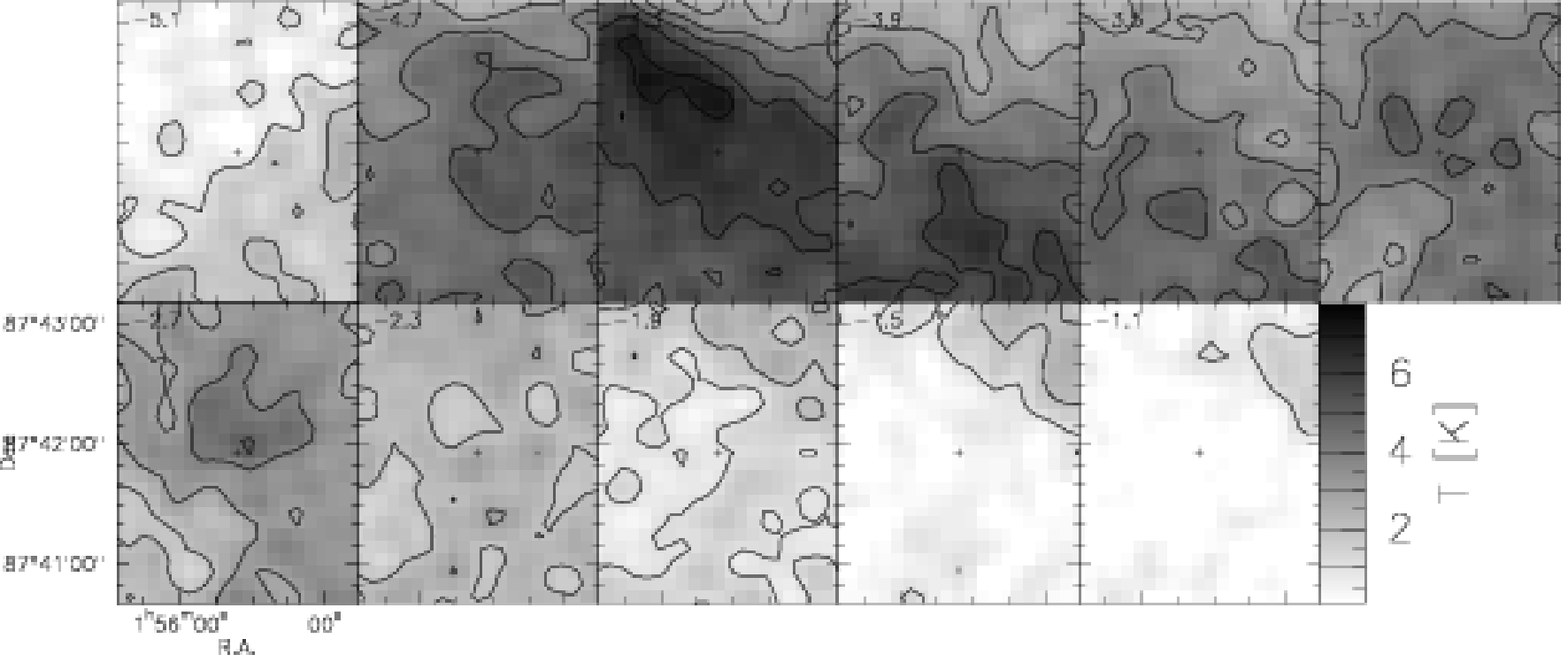}
  \caption{From top to bottom, maps of the PdBI{},
  PdBI{}+30m and 30m of \twCO(1-0) emission integrated over the same 
velocity slices of 0.3 \kms\ centered as indicated.}
  \label{fig:channel-maps}
\end{figure*}

\begin{table*}
\caption{Spatial and kinematic characteristics of the
  \twCO\ PdBI-only structures.}
\begin{tabular}{llllllllll}
\hline 
Structure & $v_{min}$ & $v_{max}$ & $\Delta
v_{1/2}$  & $T_{peak}$ & $W({\rm CO})$ & $\theta_{1/2}$ & $l_{\perp}$ $^a$  & $PA$ & $n_{\HH}^{max}$ $^b$ \\
 & \kms\ & \kms\ & \kms\ & K & \Kkms\ & arcsec & mpc & \deg\ & \cc\  \\  
\hline
1 & -5.7 & -5.2 & 0.1  & 0.6 & 0.06 & 4  & 3.0  & 109 &  1000\\
2 & -5.6 & -5.4 & 0.2  & 1.8 & 0.36 & 10 & 7.5  & 173 &  2400\\
3 & -5.2 & -4.8 & 0.2  & 2.4 & 0.48 & 9  & 6.8  & 62 &  3200\\
4 & -4.3 & -4.1 & 0.1  & 1.2 & 0.12 & 8  & 6.0  & 59  &  1000\\
5 & -3.4 & -2.6 & 0.4  & 4   & 1.6  & 12 & 9.0  & 91  &  8900\\
6 & -3.4 & -2.6 & 0.25 & 1.2 & 0.3  & 10 & 7.5  & 161 &  2000\\
7 & -3.2 & -3.0 & 0.15 & 1.2 & 0.18 & 15 & 11.3 & 173 &   800\\
8 & -1.7 & -1.3 & 0.15 & 1.2 & 0.18 & 9 & 6.8  & 59  &  1200\\
\hline
\end{tabular}
\begin{list}{}{}
\item[$^a$] projected thickness of the filamentary structures  deconvolved from beam size
\item[$^b$] upper limit because computed as $n_{\HH}=N(\HH)/l_{\perp}$
  instead of using $l_{\parallel}$ with $N(\HH)$ derived from $W({\rm
    CO})$ (see text)
\end{list}
\end{table*}

\subsection{PdBI structures: sharp edges of extended structures}

At the adopted cloud distance of $d=150$ pc, $1~''$ corresponds to
0.75 mpc or 150 AU, so that the spatial resolution of the PdBI data is
3.2 mpc or 660 AU.

The integrated emission detected with the PdBI is displayed in
Fig.~\ref{fig:integrated} (left panel), with the corresponding
signal-to-noise ratio (right panel). The integrated emission covers
most of the mosaic area.  This is no longer true when this emission is
displayed in velocity slices (Fig.~\ref{fig:channel-maps}, top
panels).  Several distinct structures are detected in addition to the
bright CO peak, at velocities [ -3.1, -2.3 ] \kms. Most are weak (the
first level in the PdBI channel maps of Fig.~\ref{fig:channel-maps} is
3$\sigma$) but they extend over many contiguous synthesized beams (10 to 30).

The PdBI data merged with the short-spacings provided by the 30m
telescope and the \twCO(1-0) emission detected by the IRAM-30m
telescope are displayed in the same velocity-slices, for comparison,
in Fig.~\ref{fig:channel-maps}, central and bottom panels
respectively.   Most of the structures seen by the PdBI lie at the
  edge in space and in velocity space of extended emission present in
  the single-dish channel maps. This property is most visible for the
two structures in the north-west of the mosaic over [-4.8, -4.4]
\kms\ and [-2, -1.2] \kms, and in the central region at $v=-2.8$ \kms.
It is even better seen by comparing the single-dish maps before and
after combination with the PdBI data. The single-dish maps are changed
in two-ways: the structures exhibit sharper, more coherent boundaries
and these boundaries extend further in velocity-space (e.g. channels
-4.7 and -2.3 \kms). In a given channel of width $\Delta v_c$, the
size of the detected structures in the CO emission $\Delta x_c$ is
inversely proportional to the velocity-shear, $ \Delta x_c = \Delta
v_c / (\partial v_{LSR}/\partial x_{pos})$.  Hence, the detection of 
small-scale structures at the edge of the velocity coverage of larger-scale
structures may be favored by an  increase of the velocity shear at
these edges.

The fact that these structures appear both in PdBI{}-only data and in
combined (PdBI{}+30m) data gives confidence in their reality,
independently of the deconvolution techniques.

In summary, the interferometer is sensitive  by construction to small-scale
  (\ie{} sharp) variations of the space-velocity CO distribution. It
  happens that the sharp structures detected by the interferometer lie
  at the edge in space and velocity of regions of shallow CO emission that extend 
over at least arcminutes, as displayed in the 30m channel maps.
The PdBI-structures are therefore the sharp edges of extended structures.

\begin{figure*}
  \centering
  \includegraphics[width=0.47\hsize{}]{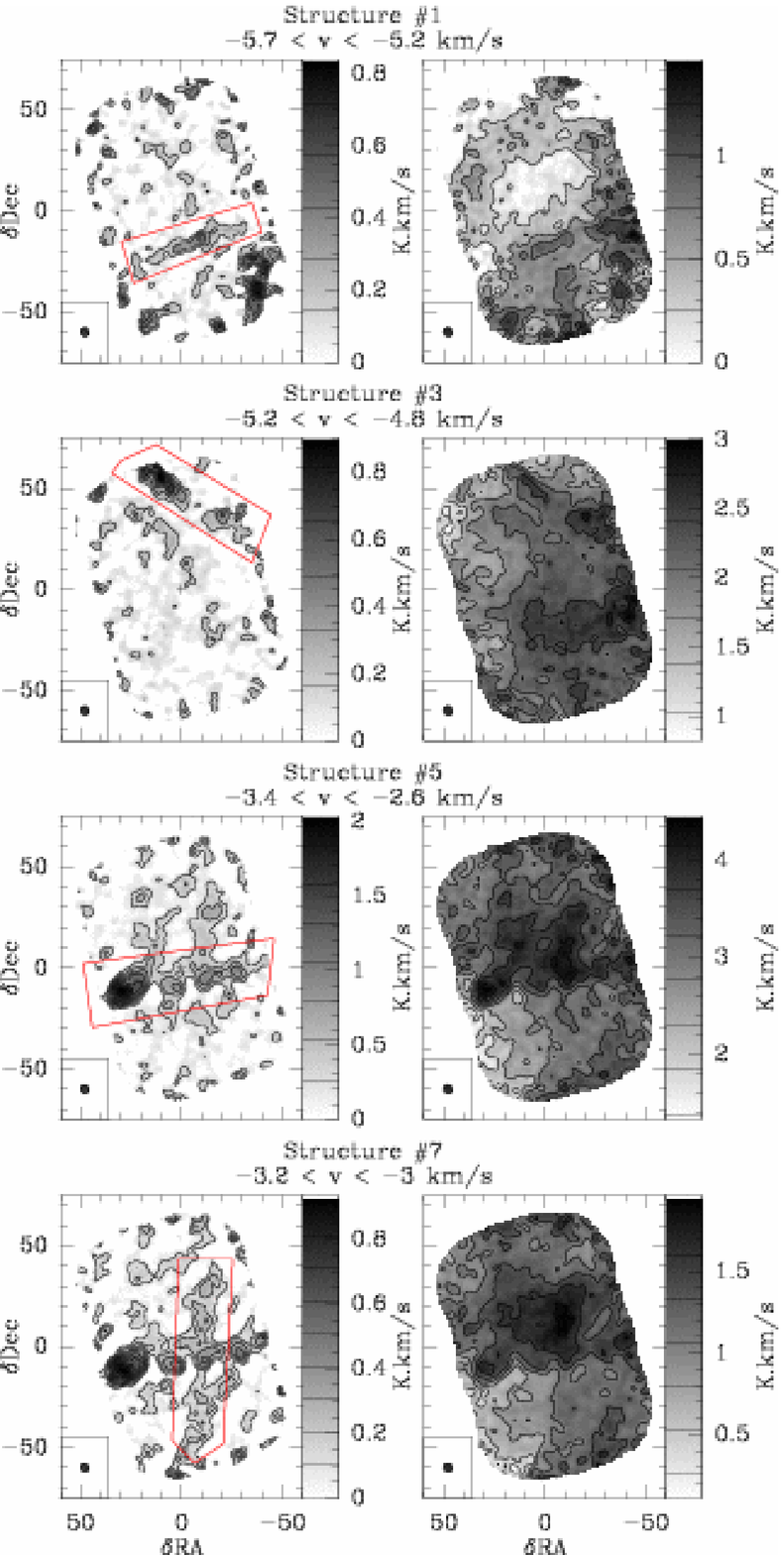}
  \hspace{0.025\hsize{}}
  \includegraphics[width=0.47\hsize{}]{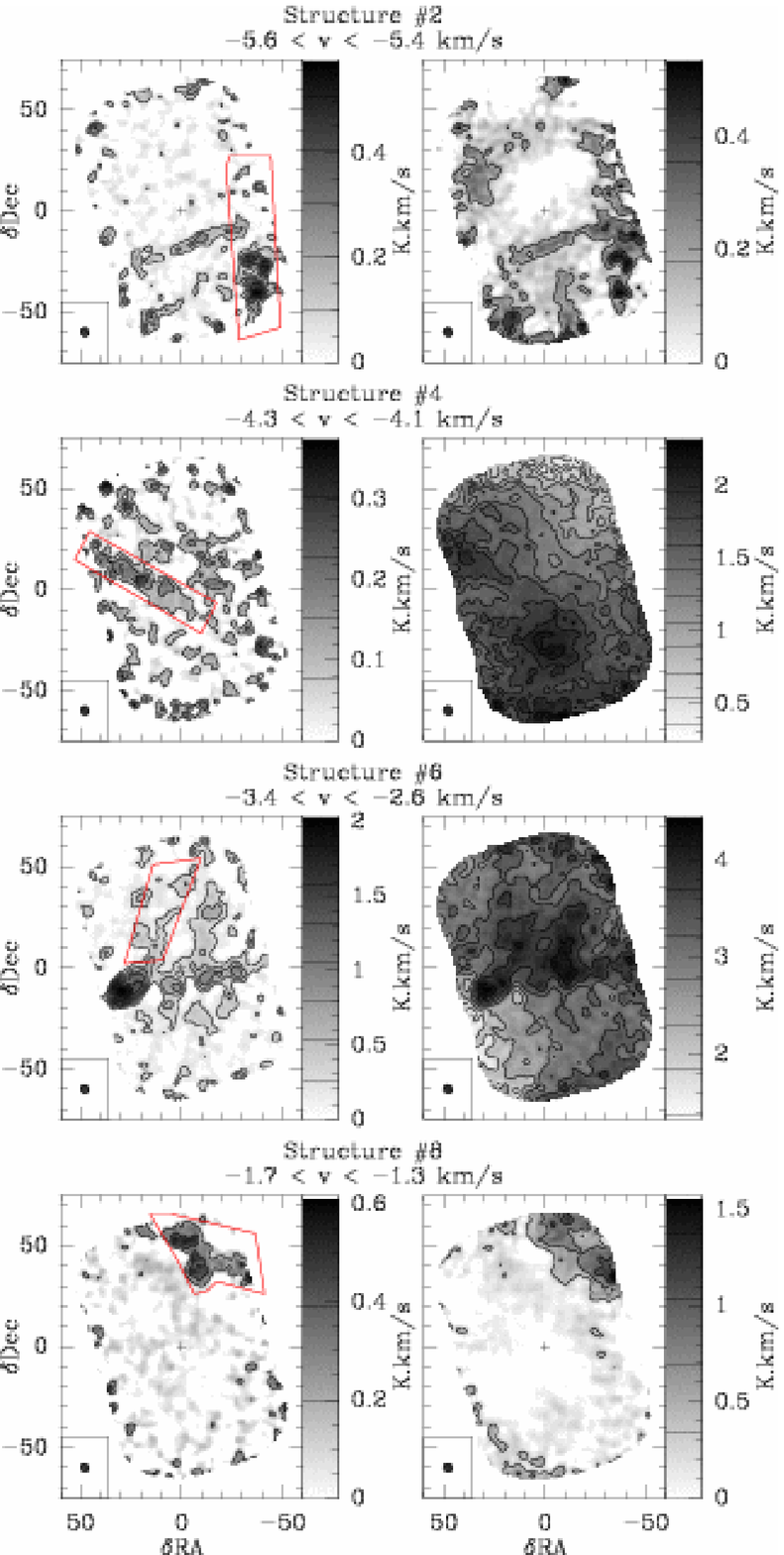}
  \caption{The 8 structures described in Table 1: {\it Left panels}:
    PdBI-only \twCO(1-0) emission integrated over the indicated
    velocity interval appropriate to each structure. {\it Right
      panels}: Same for the combined PdBI+30m mission. The polygons
    show the area over which the CO spectra of Fig.\ref{fig:spectra}
    are computed. }
  \label{fig:filaments}
\end{figure*}

\subsection{Observed characteristics of the PdBI structures}
We have identified eight structures in the space-velocity
\twCO\Jone\ PdBI data cube that are well separated from one another in
direction and in velocity. They are shown in Fig.~\ref{fig:filaments},
each drawn over its proper velocity range.
The right panels show the PdBI data
combined with 30m data over the same velocity ranges to further
illustrate that the PdBI filtering emphasizes the sharpness of the edge of
the space-velocity structures.
Fig.~\ref{fig:filaments} also shows that the single-dish structures cover a large fraction
of the mosaic area. For instance, 
in the case of structure \#1, the single-dish structure extends over 
the whole southern half of the mosaic, while for structure \#5 it almost 
covers the northern half. 

 The observed properties of the 8 PdBI structures are given in Table 2.
 The peak \twCO\Jone\ temperature is that detected by the PdBI,
 therefore the excess above the extended background, resolved out by
 the PdBI.  The size $\theta_{1/2}$ is the half-power thickness of the
 elongated structures, deconvolved from the beam size.  The projected
 thickness, in mpc, is called $l_{\perp}$ by opposition to the unknown
 depth along the line-of-sight ({\it los}), called
 $l_{\parallel}$.  The position-angle $PA$ is 
that of the direction defined, within $\pm 10\deg$, by the three brightest 
pixels of each structure.  
 When they are not aligned, as in the case of \#8, we determine a direction
with the meaning of a least-square fit. It corresponds to an average $PA$ over the detected 
structure that does not take into account the substructure visible in Fig.~\ref{fig:gradient}
for instance.
Because of their different velocity width
 and CO line temperature, the CO integrated brightness of the eight
 structures varies by a factor 25.

\begin{figure*}
  \centering
  \includegraphics[width=0.8\hsize{}]{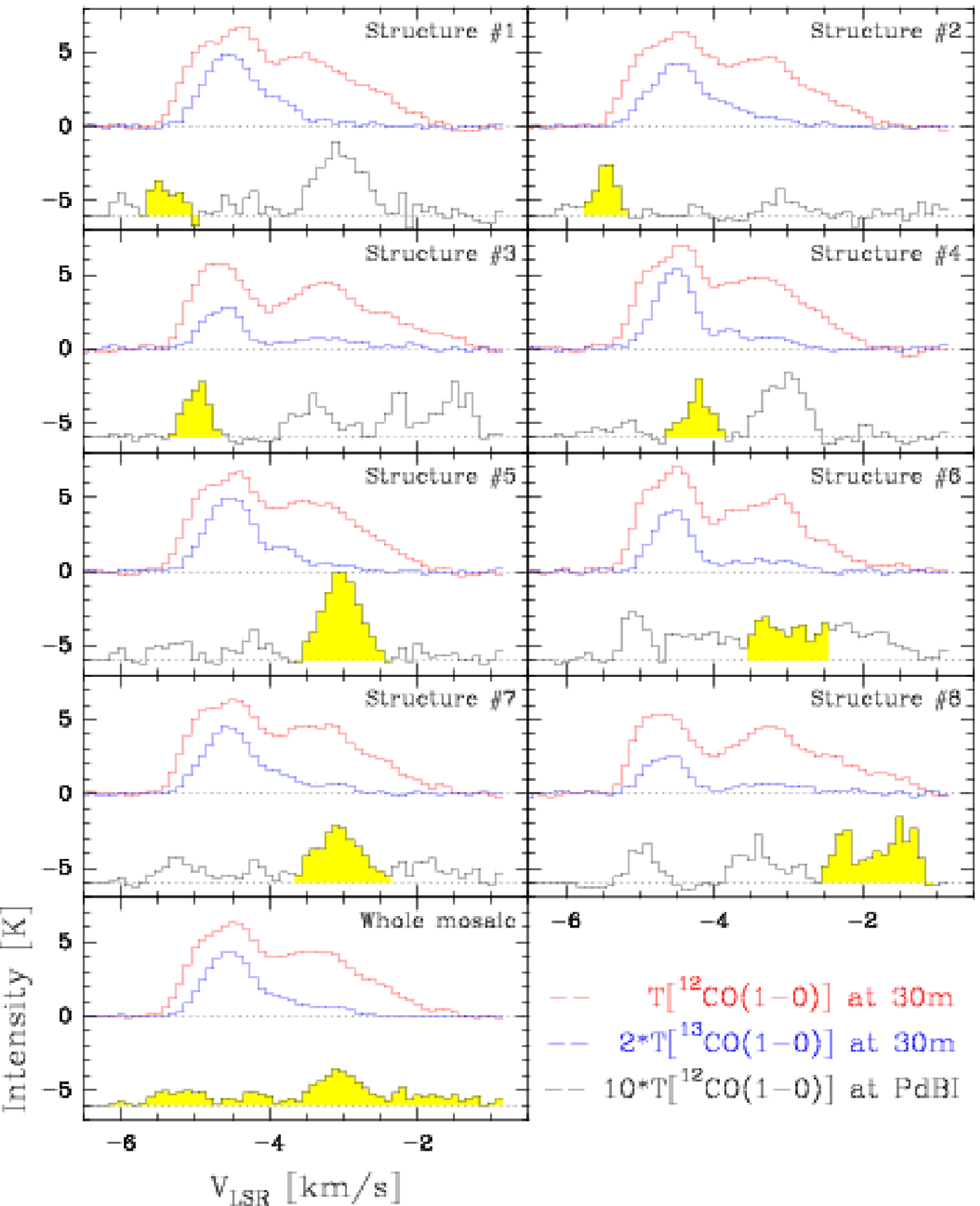}
  \caption{Comparison of spectra integrated over either the polygons
    defined in Fig.~\ref{fig:filaments} and the whole mosaic. The \twCO{} and
    \thCO{} \Jone{} single-dish spectra are shown resp. in light and darker
grey,
    while the PdBI{} only \twCO{} \Jone{} spectra is shown in black. \thCO{}
    amplitude have been multiplied by a factor of two and PdBI{} only \twCO{}
    amplitude by a factor of 10. 
    Note that \emph{i)} only a small
    fraction of the single-dish flux is recovered at PdBI{} and \emph{ii)}
    the centroid velocities of the 
 small-scale structures are all, but one, outside that of the \thCO\ peak.}
  \label{fig:spectra}
\end{figure*}

Most of the PdBI-structures are elongated and straight with different
position-angles in the sky.  Interestingly, they do not shadow each other in space
and in velocity space (\ie\ each fills only a small area of the mosaic
in a small velocity interval, and the positions and areas of the
detected structures are different). Their cumulative surface filling
factor in the mosaic field is large, $f_S \approx 0.5$
(Fig.~\ref{fig:integrated}), \ie\ $f_S = 0.6$ for the structures detected at
more than 1-sigma and $f_S = 0.3$ for 3-sigma detections.  However, the
fraction of the single-dish power (integrated over the mosaic) seen by
the PdBI in the \twCO(1-0) line is low. It depends on the velocity
interval: it varies between 2\% in the \twCO\ line-core (defined as
the velocity range, [-5.0, -3.5]\kms, over which the single-dish
\thCO/\twCO\ is the largest, see F98), and 6\% in the line-wings.
Fig.~\ref{fig:spectra} displays the emission profile of the 8
PdBI-structures with the single-dish \twCO\ and \thCO\Jone\ emissions
over the same area (defined by the polygons of
Fig.\ref{fig:filaments}).

Last, the PdBI-structures cover the full velocity range of the
single-dish CO line (see bottom panel of Fig.~\ref{fig:spectra})
including the far line-wings (\eg\ structure \#2 at -5.5 \kms).  Note
however that the spectrum integrated over the whole mosaic peaks at -3
\kms, in the wing of the single-dish \twCO\ line while its minimum, around -4.5 \kms,  
coincides with the peak of the single-dish \thCO\ line (\ie\ line core).
 The broad velocity distribution of the
PdBI-structures within the single-dish line coverage ensures that they
are not artefacts of radiative transfer. If they were, they would
appear preferentially at extreme velocities because CO photons escape probability
is larger there.
There may be a small effect since the power
fraction in the line-wings is slightly larger than in the line-core,
but these fractions are a few percent in each case.  The structures
found are therefore real edges in space and velocity-space of larger
structures.

In this respect, it is interesting to place each PdBI-structure in its
\twCO(1-0) larger-scale environment at the appropriate velocity
(Fig.~\ref{fig:environment:12co10}).  The PdBI-structures, marked as
 polygons, lie at the edge of structures that
extend  beyond the field of the mosaic, up to $\sim 300~''$ or 0.2 pc.  
In the case of structures
\#3, \#4 and \#5, the orientation of the edges of the large-scale
patterns is more visible in the \thCO(1-0) maps
(Fig.~\ref{fig:environment:13co10}), likely because of the \twCO(1-0)
optical depth.  This coincidence strongly suggests that the
orientation of the PdBI-structures is not only real but also rooted in the
larger-scale environment.
 
\subsection{Pairs of parallel structures }

\begin{figure}
  \centering
  \includegraphics[height=\hsize{},angle=270]{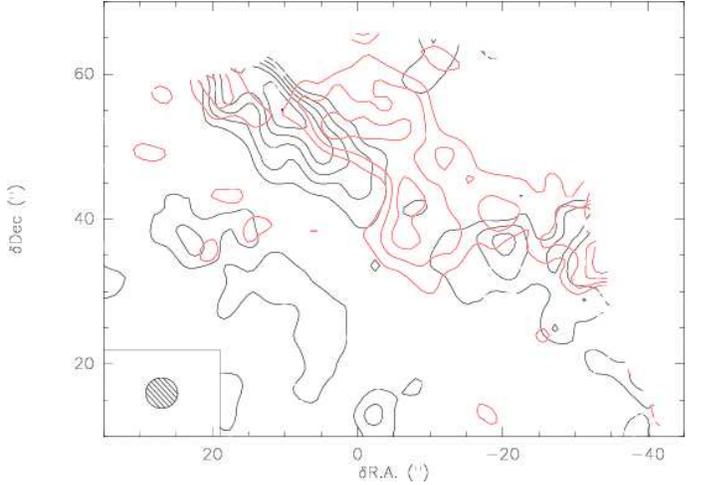}
  \caption{Structures \#3 at -5 \kms\ and \#8 at -1.5 \kms\ overplotted (resp. black and red
    contours) to display their close spatial correlation.}
  \label{fig:gradient}
\end{figure}

One of the most challenging finding of this study is the fact that
among the eight elongated PdBI-structures, six form 3 close pairs
(separated by less than $20~''$ in projection) of structures
parallel within $\pm 10\deg$ (Table 2 and Fig.~\ref{fig:filaments}).
These are the pairs of structures [\#3, \#8], [\#1, \#5] and [\#6,
  \#7].  The average position-angles of each pair $\overline{PA}= $ 60, 100 and
168$\deg$ are all different. Since the structures (at least in the two
first pairs) are at different velocities, they are not due to
artefacts of the deconvolution process.

The probability of a chance association of these three pairs in the
field of the mosaic is estimated to be at most $4 \times 10^{-9}$. 
It is the cube of the probability of having one close pair of parallel structures. 
The latter is the product of the probability, equal to $5.4\times10^{-3}$, 
that two, out of eight,
randomly oriented straight structures be aligned within
$\pm$10\deg\ of each other (\ie\ be together in a solid angle $\Delta
\Omega= 0.1$ sr), by that (ranging between 0.2 and 0.3
depending on the orientation of the pair) to be separated in projection by less
than $20~''$ in a mosaic of $1' \times 2'$. The probability of a chance association
is only slightly underestimated if one considers the structure \#8 that is not straight, strictly 
speaking.

Since the probability of a chance association of the observed pairs is
so low, we infer that the pairs are real associations.  This physical
connexion is supported by the detail of the spatial distribution of
the \twCO\ emission integrated over the two velocity ranges of
structures \#3 and \#8 in Fig.~\ref{fig:gradient}: the hole, in the
low-velocity emission is filled in by high-velocity emission, while a
common average orientation exists over $\sim 1 '$ for the pair.

\begin{figure*}
  \centering
  \includegraphics[width=0.3\hsize{},angle=-90]{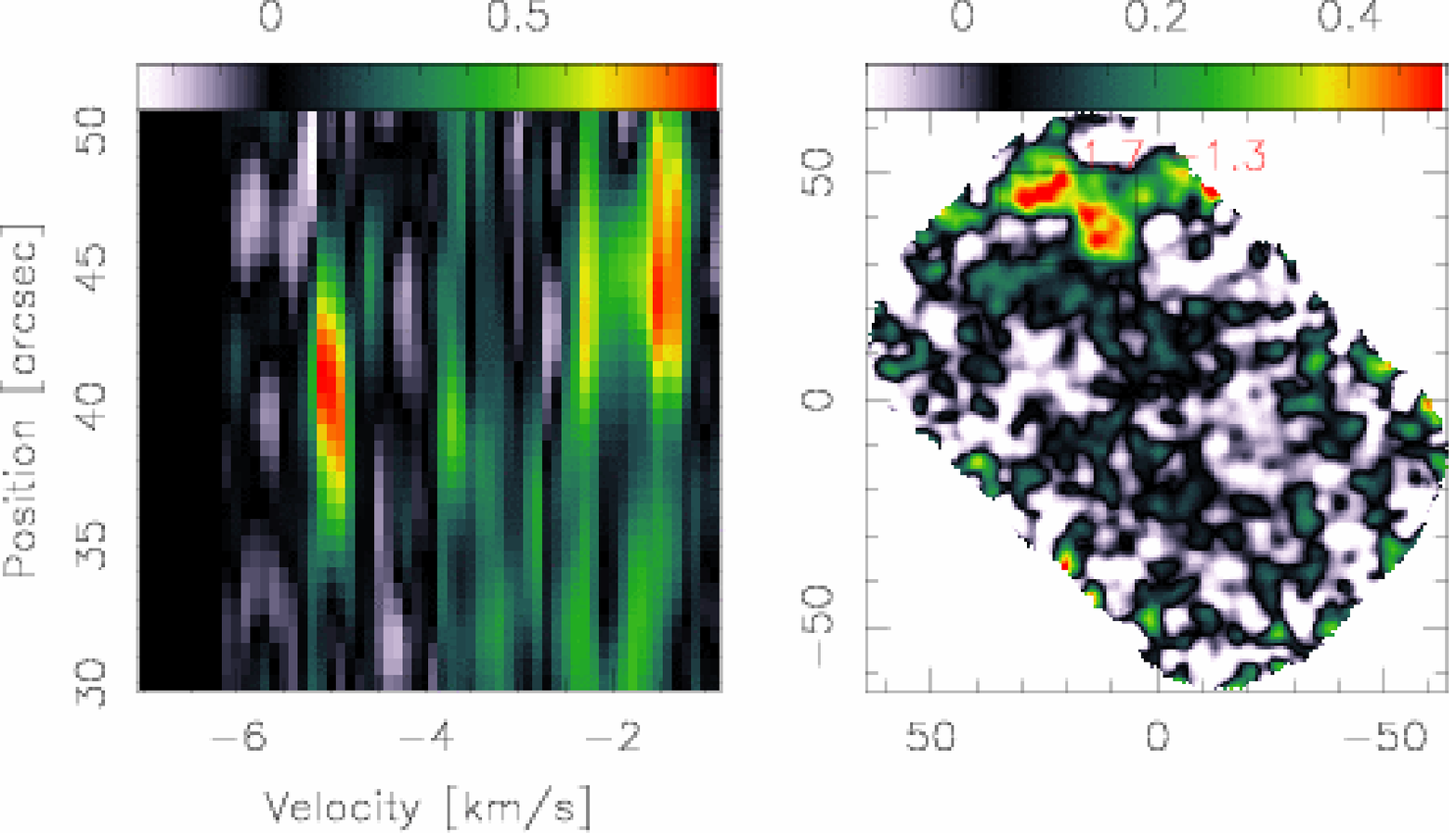}
  \includegraphics[width=0.3\hsize{},angle=-90]{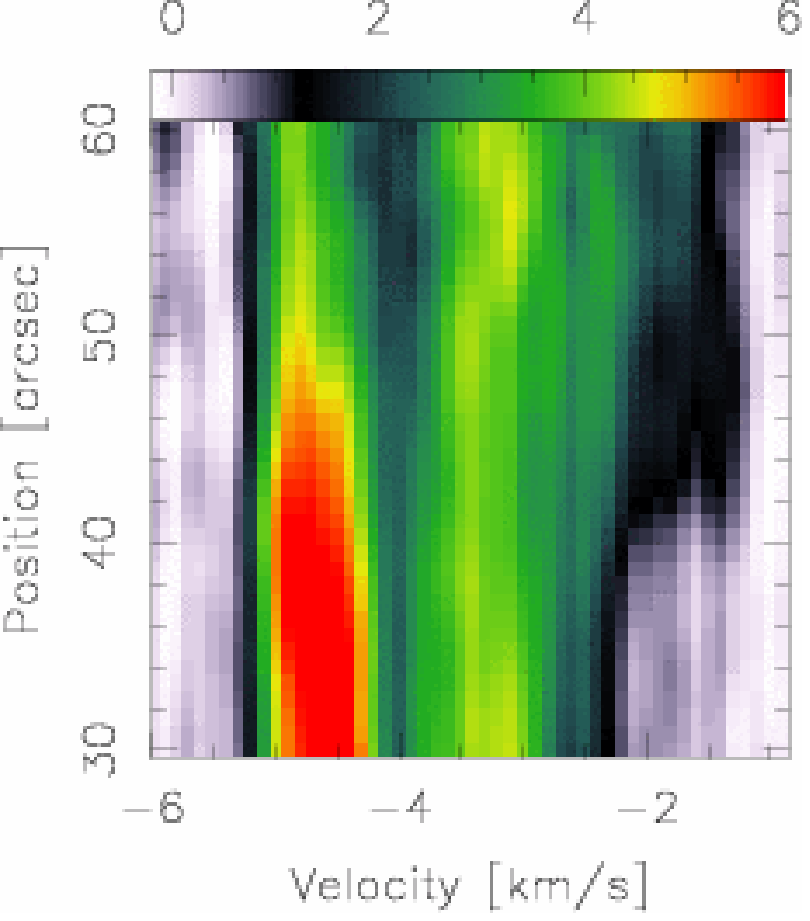}
  \caption{Position-velocity diagrams across the pair of structures [\#3,\#8]. {\it Left:} 
Cut across
the PdBI-only map. {\it Center:} Rotated PdBI channel-map [-1.7, -1.3] \kms, showing the 
direction and distance (horizontal size of the box) over which
the CO emission is averaged for the cut. The cut runs from the southern to northern edge 
of the box.  
{\it Right:} Same across the PdBI+30m data cube.}
  \label{fig:pvcut}
\end{figure*}

Two position-velocity cuts (Fig.~\ref{fig:pvcut}) across the pair [\#3,\#8]
further illustrate what is meant by  sharp edges and real association. 
The cut across the PdBI-only data cube (left panel) 
shows two CO peaks centered at offset positions $40~''$ (resp. $46~''$) and velocities 
-4.9 \kms\ (resp. -1.6 \kms) for the low- and high-velocity component respectively. 
%With a HPW of 7 and 9 ~arcsec respectively, they are resolved by the observations.  
 These resolved peaks are located exactly  at the terminal pixels  of the 
larger-scale structures visible at the same velocities in the cut across the 
PdBI+30m data cube (right panel). On this cut, the low-velocity component may be followed 
over all offsets below $\approx 46~''$, while the high-velocity component is 
visible at all offsets above $\approx 40~''$. This cut also illustrates 
a clear difference between the two velocity components:
the velocity of structure \#3 
(peak at -5 \kms\ in the PdBI spectrum of Fig.~\ref{fig:spectra}) 
falls within the velocity coverage 
of the bright extended gas (\twCO\ and \thCO\ line core in the single-dish spectra) while 
that of structure \#8 (peak at -1.5 \kms\ in the PdBI spectrum of Fig.~\ref{fig:spectra}) is 
not blended with any other emission in that extreme velocity range
and appears as a weak emission in the single-dish spectrum (\ie\ a line-wing).
Such blendings in space and velocity projections with   
extended components resolved out by the PdBI observations (Fig.~\ref{fig:pvcut}, right panel)
explain why such pairs of structures are so difficult to recognize in single-dish observations or 
low sensitivity interferometric observations.

The PdBI-structures cannot therefore be understood as isolated
entities.  Not only  are they the sharp edges of larger CO-structures seen in the
single-dish maps  but also 6 out of 8 of these edges are paired. In the following, we will call 
the  CO extended structures bounded by sharp edges either Sharp-Edged Extended
Structures (SEES) or  Sharp-Edged Extended Double Structures (SEEDS) when they  belong to a pair, 
to emphasize this essential property.
   
\subsection{Velocity shears}

The pairs being real associations, we     
ascribe a velocity-shear to each of them. The projected separation 
$\delta l_{\perp}$ and velocity difference $\delta v_{LSR}$ between the
low- and high-velocity components of each pair provide a measure of the velocity-shear  
 $\delta v_{LSR}/ \delta l_{\perp}$. We cannot determine whether this measure
is a lower or upper limit of the true velocity-shears because of the projection effects: both
the separation measured in the {\it pos}  and the velocity difference are lower limits.

The results are given in Table 4.
The method used is illustrated  in Fig.~\ref{fig:pvcut} (left panel) for the 
pair [\#3,\#8]: the projected separation between the low- and high-velocity components is 
$6~''$ or 4.5 mpc while the velocity separation is 3.5 \kms, hence 
a velocity-shear of 777 \kmspc,  the largest ever measured in CO emission in a
molecular cloud devoided of star formation activity. 
These values correspond to an average 
over several positions along the shear direction, including those where
the two velocity components partially overlap. The separation is therefore 
slightly underestimated by the averaging.  
 Note that one pair only, [\#6, \#7], has a very small velocity-shear, 
probably because, in that case, the two velocity components
 involved in the shear are mostly in the {\it pos}.
A rate-of-strain, defined as 
$a= {1 \over 2} \delta v_{LSR}/\delta l_{\perp}$, and 
timescale $\tau=a^{-1}$, are also  given
to help comparison with chemical models (Section 7).  The large 
observed velocity-shears translate into timescales as short
as a few 10$^3$ yr, if the Lagrangian and Eulerian views 
of the fluid  can be exchanged  \citep[see][]{mordant02} . 

\begin{table*}
\caption{Spatial and kinematic characteristics of the
 three pairs of parallel PdBI-only structures.}
\begin{tabular}{lllllllll}
\hline 
Pair & $v_{1}$ & $v_{2}$ & $\delta v_{LSR}$  & $\delta l_{\perp}$ $^a$ & $\delta v/\delta l_{\perp}$ & $a$ $^b$ & $\tau$ $^c$ & $l$ $^d$    \\
\hline
  & \kms\ & \kms\ & \kms\ & mpc & \kmspc\ & s$^{-1}$ & yr & mpc  \\
\hline
\#3, \#8 & -5.0 & -1.5 &  3.5 & 4.5  & 777   & 1.3$\times 10^{-11}$ & 2.5$\times 10^3$ & 45  \\
\#1, \#5 & -5.4 & -3.0 &  2.4 & 9.0 & 267 & 4.5$\times 10^{-12}$ & 4$\times 10^3$ & 45   \\
\#6, \#7 & -3.0 & -3.1 &  0.1 & 16 & 6 & $10^{-13}$ & 3$\times 10^5$ & 40   \\
\hline
\end{tabular}
\begin{list}{}{}
\item[$^a$] averaged separation between the PdBI CO peaks
\item[$^b$] $a= {1 \over 2} \delta v/\delta l_{\perp}$
\item[$^c$] $\tau=a^{-1}$
\item[$^d$] length over which the structures are parallel within $\pm 10\deg$.
\end{list}
\end{table*}

\subsection{The SEE(D)S are layers of CO emission}

The small-scale structures detected by the PdBI have properties never
seen before because the present observations are most sensitive and
the field of view is large in comparison to the resolution: (1) they
are not clumps, but elongated structures, only bounded by the limited
size of the mosaic, (2) they all mark a sharp fall-off of the CO
emission in selected velocity ranges: they are not isolated filaments,
but the sharp edges (3 to 11 mpc in projection), simultaneously in
space and velocity-space, of larger structures, the SEE(D)S, extending
beyond the mosaic ($l>0.2$ pc), (3) six of these form three pairs of
parallel structures at different velocities, with a small projected
separation and the velocity-shears estimated for two of these pairs,
several 100 \kmspc, are the largest ever measured in non-star forming
clouds.

  If the SEE(D)S were  CO-emitting volumes (\ie\ 3-dimensional
   structures in space) of characteristic dimension $l$, their edges would be
  surfaces commensurate with $l^2$.  In projection, these edges
  would appear as surfaces, also commensurate with $l^2$ for a random
  viewing angle.  Only if these surfaces were plane and viewed edge-on
  (within $\pm 5$ deg for a projected size less than one tenth of
  their real size) would  these edges appear as thin elongated
  structures.   We rule this out on statistical grounds: the mere
    fact that we detect 8 sharp CO-edges in the small field-of-view of
    the PdBI observations suggests that it is not a rare configuration
    and that the eight sharp CO-edges are seen from random viewing
    angles. We thus infer that the SEE(D)S are  CO-layers, 
    rather than volumes and that their thickness is  $\sim 10$ mpc or
    less, on the order of the width of the PdBI-structures.

  This statistical argument is reinforced by the presence of pairs. The SEEDS
  are structures that have sharp edges with only small or null overlaps. If
  their interface were 2-dimensional (\ie{} if the SEEDS were volumes), the
  small overlap would occur only for an edge-on viewing, an unlikely case.
  Their interface is  therefore 1-dimensional rather than 2-dimensional and
  the SEEDS are  layers of CO emission.  This ensures that
  under any viewing angle the two extended velocity components are detected
  with only a narrow or null  spatial overlap in projection.
  The SEEDS could still be 
be 3-dimensional pure velocity-structures, where large velocity-shears produce sharp 
edges in  channel maps of finite spectral resolution (see Section 3.1). 
However, with the same statistical argument as above, concerning density structures, 
we rule out the possibility that the SEEDS be {\emph 3-dimensional} velocity structures. 
These must be CO layers. 

In summary,  the sharpness of the edges of the SEE(D)S, associated
with the fact that we detect 8 cases in the mosaic and three close-pairs 
that do not overlap in space,  implies that 
the SEE(D)S are thin layers of  CO emission rather than volumes.  

\section{Gas density of the PdBI-structures}

\subsection{Estimates from CO line emission}

Because of the elongated shape of most of the structures
and the fact that they are edges of more extended emission, 
we have not tried to decompose the observed emission using clump
finding algorithms such as GAUSSCLUMP \citep{stutgust90}. We
estimate below the gas density in these structures in two independent
ways and compare the results to those inferred from the dust continuum
emission.
 
First, we compute upper limits of  
the \HH\ densities (Table 2) by 
adopting the CO-to-\HH\ conversion factor 
$X= 1.56\pm0.05 \times 10^{20}$ \cq\ (\Kkms)$^{-1}$ \citep{hunter97}
so that $n_{\HH} = 5\times10^4 \,\cc\ W({\rm CO})/l_{mpc}$, for a 
{\it los} depth equal to the projected thickness $l_{\perp}$. 
Since we are observing edges of layers (see section 3.2),
 the inferred densities are overestimated by the unknown 
factor $l_{\parallel}/l_{\perp}$.
 The upper limits of the 
\HH\ densities derived from the
galactic CO to \HH\ conversion factor (Table 2) vary by a factor 10 only.

Alternatively, one may use a LVG analysis to estimate the gas
properties in these structures. Two assumptions are made: (1) the CO
emission is not beam-diluted and (2) the excitation is assumed to be
the same as measured in the same field with the IRAM-30m so that we
adopt the line ratio $R(2-1/1-0) =0.7\pm$ 0.1 \citep{falgarone98kp,
hily08}.  This value may be representative of
the excitation of translucent molecular gas because the same line
ratio is found in different observations sampling a similar kind of
molecular gas \citep{pety08}.  Under these conditions, the CO
column densities per unit velocity are very well determined for all
line temperatures.  They are given in Table 4 for the brightest,
weakest, and most common CO peak temperature observed. The inferred CO
column densities differ by only a factor 10 to 16 between the brightest
and weakest structure.

Table 4 also gives the range of gas kinetic temperatures and
associated range of densities, thermal pressures $P_{th}/k$ and CO
abundances of possible solutions.  The range of temperatures is
bounded towards high values by the thermal width of the CO lines
($T_k<250$ K for the broadest line, $<35$K for the narrowest).
Solutions colder than 10K are unlikely because the gas is poorly
shielded from the ambient ISRF. The CO optical depth is therefore
smaller than a few, in agreement with the results of Paper II.
A similar conclusion has been
derived by \cite{heithausen06} after he failed to detect the \thCO\ and
\CeiO\Jone\ line with the PdBI in a nearby small-area
 molecular structure (SAMS) field.

Each set of \HH\ density and kinetic temperature in the LVG solutions,
corresponds to a product $X({\rm CO)} l_{\parallel}$ where $X({\rm
CO)}$ is the CO abundance relative to \HH.
The range of CO
abundances inferred from the LVG computations are given in Table 2 for
$l_{\parallel}=l_{\perp}$. They may be overestimated by the unknown ratio
$l_{\parallel}/l_{\perp}$.

\begin{table*}
\caption{Results of LVG radiative
transfer calculations for representative observed values (see Table 2)}  
\begin{tabular}{lllllllll}
\hline
   & $T_{peak}$  & $N({\rm CO})/\Delta v$ $^a$ & $\Delta v$ & $N({\rm CO})$ &
$T_k$ range & $n_{\HH}$ range $^b$ & $P_{th}/k$ range $^b$ & $X({\rm CO})$ range $^b$\\ 
\hline 
  & K & \cq/\kms\ & \kms\ & \cq\ & K  &\cc\ & K \cc\  & $\times 10^{-5}$   \\ 
\hline
brightest & 4 & 3 -- 4 $\times 10^{15}$ & 0.4 & 1.2 -- 1.6 $\times 10^{15}$ 
& 10 -- 200 & 3$\times 10^3$ -- 250 & 3 -- 5 $\times 10^4$ & 2--20\\
most common  & 1.2 &  $1.5 \times 10^{15}$ &  0.2 & $3 \times 10^{14}$ & 
10 -- 140 & 8$\times 10^3$ -- 300 & 8 -- 4 $\times 10^4$ & 0.16--4\\
weakest & 0.6 &$1.0 \times 10^{15}$ & 0.1 & $1.0 \times 10^{14}$ & 7 -- 35 & 
1$\times 10^4$ -- 800 & 7 -- 3 $\times 10^4$ & 0.11 - 1.4 \\ 
\hline
\end{tabular}
\begin{list}{}{}
\item[$^a$] assuming $R(2-1/1-0)=0.7\pm 0.1$ 
\item[$^b$] the LHS (resp. RHS) values correspond to the lowest (resp. highest) gas temperature 
\end{list}
\end{table*}

Table 4 shows the range of possible \HH\ densities derived from the
LVG analysis for gas temperatures between  10K and 200K at most.
The comparison of these values with the upper limits 
 inferred from the CO-to-\HH\ conversion
factor (Table 2) provides narrower \HH\ density ranges,  $n_{\HH}= 800$ to 
$10^3$ \cc, and 300 to $2 \times 10^3$ \cc, for the weakest 
and most common structures respectively. In 
spite of all the uncertainties, the two 
methods infer consistent \HH\ densities that do not exceed $3 \times 10^3$\cc.
Moreover, whether the gas is cold or warm, its thermal pressure is
about the same, within a factor of a few, and is in harmony with that
inferred from carbon line observations in the local ISM that has an 
average of $P_{th}/k \sim 3 \times 10^3$
K \cc\ with fluctuations up to $\sim 10^5$ K \cc\ \citep{jenkins07}.

\subsection{Estimates from the dust continuum emission}

In addition to \twCO\ lines, we have detected continuum emission.
This emission is close to the noise level, except for the large bright
spot associated to the \twCO\ peak of emission.  A comparison of the
continuum emission with the CO contour levels
(Fig.~\ref{fig:integrated}) suggests that the elongated feature of
continuum emission in the north-western corner of the mosaic is also
real.

On the basis of the coincidence of the peaks of the \twCO\ and
continuum emission in the mosaic, we ascribe the continuum emission to
thermal dust emission. The average continuum brightness over the
\twCO\ peak (Fig.~\ref{fig:integrated}) is 2$\pm 1$ mJy/beam, hence
$I_{cont}= 1 \pm 0.5 \times 10^{-20}$ \ecqs\ Hz$^{-1}$.  The dust
opacity, $\tau_d= I_{cont}/B_\nu(T_d)$, depends on the dust
temperature.  We adopt a dust emissivity $\tau_d= 8.7 \times 10^{-26}
(\lambda / 250 \mu{\rm m})^{-2} \, N_H$, deduced from COBE data for
dust heated by the ambient interstellar radiation field (ISRF)
\citep{lagache99}. For $T_d=10$ K and $\nu=115$GHz, we find
$\tau_d=3.2\times 10^{-7}$ and $N_H=4.0 \times 10^{20}$ \cq\ across the
peak. For $T_d=17$ K, the most plausible value in translucent gas of
the Solar Neighborhood, this value would be lower by a factor 2 and
for $T_d=8$ K it would be larger by 40\%, providing the range $N_H=2.0
\times 10^{20}$ to $5.6\times 10^{20}$ \cq\ for high and low dust
temperatures, respectively.  This estimate of $N_H$ may be compared to
that inferred from \twCO\ in the previous section. If we allow for a
column density of atomic hydrogen comparable to that of \HH, as found
in the Polaris Flare by \cite{heithadd90}, the total H
column density inferred from CO for the peak of structure \#5 ranges
between $N_H=3 N(\HH)=2 \times 10^{19}$ to $2.4\times 10^{20}$
\cq\ for the warm and cold solutions respectively (assuming a size of
9mpc).  The two ranges of values would overlap for a depth a few times
larger than the observed projected size, allowing for warm and
moderate density solutions where the dust temperature is lower than
that of the gas.  Given the many uncertainties in the different steps
(the questionable validity of the CO line analysis, the knowledge of dust
emissivity and temperature), the consistency between these two
independent estimates is encouraging and we are confident that we have
detected the dust thermal emission of the brightest small-scale
structure and that its \HH\ density is not higher than a few
10$^3$\cc.

Aside from the peak, the rest of the structures have column densities
a few times smaller and their dust continuum emission is expected to
lie closer to the noise level. In addition, the surface filling factor
of the \twCO\ structures being large in the central area of the
mosaic, the PdBI visibility of the continuum emission of individual
structures is expected to be highly reduced compared to that of the
line which takes advantage of velocity-space.  This may be the reason
that the continuum emission and the \twCO\ emission do not coincide
elsewhere: the continuum emission is more heavily filtered out by the
interferometer than the \twCO\ emission.

\begin{figure}
  \centering
  \includegraphics[width=\hsize{}]{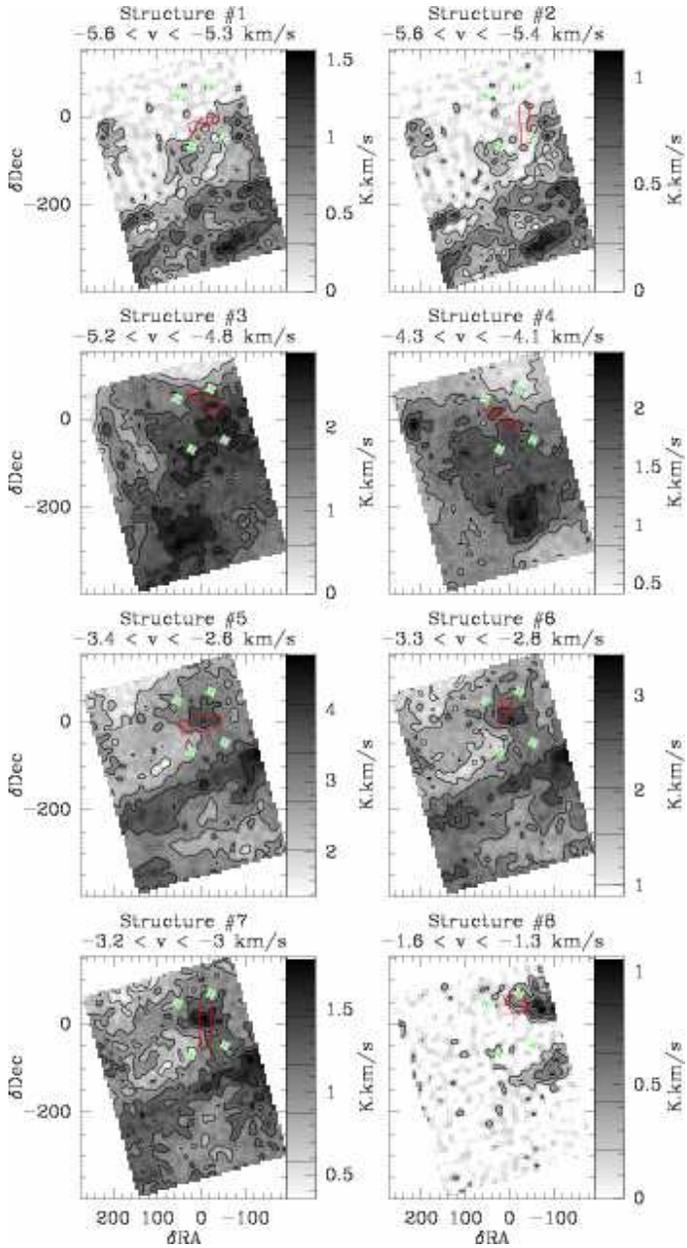}
  \caption{Integrated maps of the same velocity range as that defined
    in Fig.~\ref{fig:filaments} of the \twCO{} \Jone{} map observed at
    the IRAM-30m. Green crosses delimit the mosaic position and the
    red polygon as defined on Fig.~\ref{fig:filaments} shows the
    position of the elongated structures detected at PdBI{}.}
  \label{fig:environment:12co10}
\end{figure}

\begin{figure}
  \centering
  \includegraphics[width=\hsize{}]{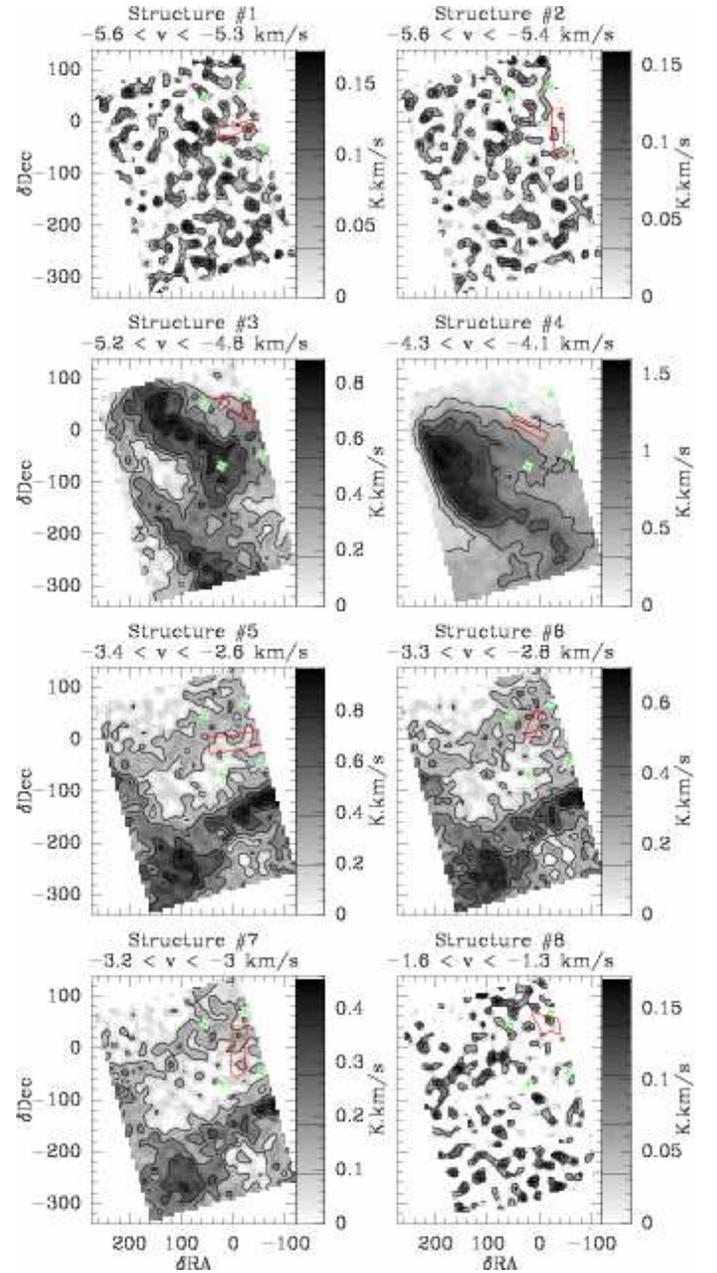}
  \caption{Same as Fig.~\ref{fig:environment:12co10} except that the single
  dish map is the one of \twCO{} \Jone{}.}
  \label{fig:environment:13co10}
\end{figure}

\section{What are the SEE(D)S?}
 
\subsection{Manifestations of the small-scale intermittency of turbulence}

The two largest velocity-shears given in Table 3 are more than two
orders of magnitude larger (within the uncertainties due to
projections) than the average value of 1~\kmspc\ estimated on the
parsec scale in molecular clouds \citep{goldsmith1985}. The velocity
field in these two SEEDS therefore significantly departs from
predictions based on scaling laws obtained from \twCO(1-0) in molecular clouds, such as
that shown in Fig.\ref{fig:vntr}. 
In spite of a significant scatter of the data points, a power law $\delta
v_l \propto l^{1/2}$ characterizes the increase of the velocity
fluctuations with the size-scale $l$, at least above $\sim 1$~pc.  Below that scale-length,
the scatter increases and a slope 1/3 would not be inconsistent with the data.
According to the former scaling, the
velocity-shear should increase as $l^{-1/2}$, therefore by no more
than $140^{1/2}= 12$ between 1~pc and 7~mpc. If the other scaling is adopted, 
this factor becomes $140^{2/3}= 26$. Now, the observed shears
increase by more than two orders of magnitude between these two scales.
This is conspicuous on Fig.\ref{fig:vntr} where the 8 PdBI-structures of Table 2 are 
plotted either individually  or as pairs (\ie\ as they 
would be characterized if the spatial resolution were poorer and individual structures 
were not isolated in space, providing for instance a linewidth $\Delta v_{1/2}=3.5$ \kms\ 
and a size $l_{\perp} \sim 7$ mpc  for the pair [\#3,\#8]).

This result  has to be put in  the broader perspective described in Section 1.
The statistical analysis of the velocity field of this high latitude
cloud (Paper III, HF09) shows that the {\it pdf} of the \twCO\ line-centroid
velocity increments increasingly departs from Gaussian as
  the lags over which the increments are measured decrease.  The
locus of the positions that populate the {\it pdf} non-Gaussian wings  forms
elongated and thin ($\sim 0.03$ \pc) structures that have a remarkable
coherence, up to more than a parsec. HF09 propose, on this statistical
basis, but also because of their thermal and chemical properties given in Section 1, 
that these structures trace  the intermittency of turbulent dissipation 
 in the field.  The pair of structures [\#3,\#8] 
 belongs to that locus of positions (see their Fig. 3). The extremely
large velocity-shears measured in that small field are not just
exceptional values: they have to be understood as a manifestation of
the small-scale intermittency of interstellar turbulence, as studied
on statistical grounds in a much larger field.

\subsection{The emergence of CO-rich gas}

The PdBI-structures mark sharp edges in the \twCO\ emission. 
As discussed in Section 3.3 and illustrated in Fig.~\ref{fig:pvcut}, 
the CO emission of space-velocity structures extending over 
arcminutes (the SEE(D)S) 
drops  below the detection level over $4.3~''$ (the resolution).
Therefore, several questions arise: what is the nature of the undetected gas that 
provides the continuity of the flow? Is it undetected because its density is too low to excite the 
$J=1-0$ transition of \twCO? Or is it dense enough but with too low a CO abundance?
For simplicity, in the following discussion, ``CO-rich'' qualifies  the gas with 
$X({\rm CO}) > 10^{-6}$, the CO abundance of the weakest detected structure (Table 4), and 
``CO-poor'' the gas with a lower CO abundance. 

Acording to LVG calculations, the 3$\sigma$ detection limit of our observations allows us to 
detect CO column densities as low as $N({\rm CO}) \sim$ a few $10^{14}$ \cq\ in gas 
as diluted as $n_{\rm H} \sim 50$\cc, at any temperature, and for a velocity dispersion of 0.2 \kms, 
characteristic of the structures found. This detection limit is very low.
Therefore, if the undetected gas on the other side of the edge 
is CO-rich (with a total hydrogen column density comparable to that of the detected part), 
it has to be at a density lower than  $n_{\rm H}
\sim 50$\cc, not to excite the \twCO\Jone\ transition at a detectable level.
We rule out this possibility because 
this density is that of the CNM and it is unlikely that gas at that density be CO-rich 
(see also the models of \cite{pety08}).  
 
The alternative is that the undetected gas is  CO-poor
and that it is not its low density but its 
low CO abundance that makes it escape detection in \twCO\Jone.
Given the sharpness of the edges of the SEE(D)S, between 3 and 11 mpc (Table 2), 
the process responsible for this transition has to be able to 
generate a significant  CO enrichment over that small scale. 

In the above, we rule out the possibility that the sharp edges (\ie\ the PdBI-structures)
mark photodissociation fronts, because the orientations of such fronts
would not be randomly distributed, as is observed. Moreover, there is
no source of UV photons in that high-latitude cloud and the radiation
field there is the ambient galactic ISRF. Photodissociation fronts
would not have different orientations depending on gas velocities
varying by only a few \kms.
The sharp edges are not either folds in  layers of CO emission because
those who belong to SEES (single structures) lack the second part of the layer, 
and those who belong to SEEDS have the two parts at different velocities.

We thus infer that the SEE(D)S are the outcome of a dynamical process,
that involves large velocity-shears, and takes place in a gas
undetected in \twCO(1-0) emission,  because it is CO-poor, not
  because it is too diluted. This gas may be the CNM and the 
  dynamical process has to be able to enrich the CNM in CO molecules
  within a few $10^3$ yr and over a few milliparsec. 

\section{Comparison with other data sets}

\begin{figure}
  \centering
  \includegraphics[width=0.7\hsize{},angle=-90]{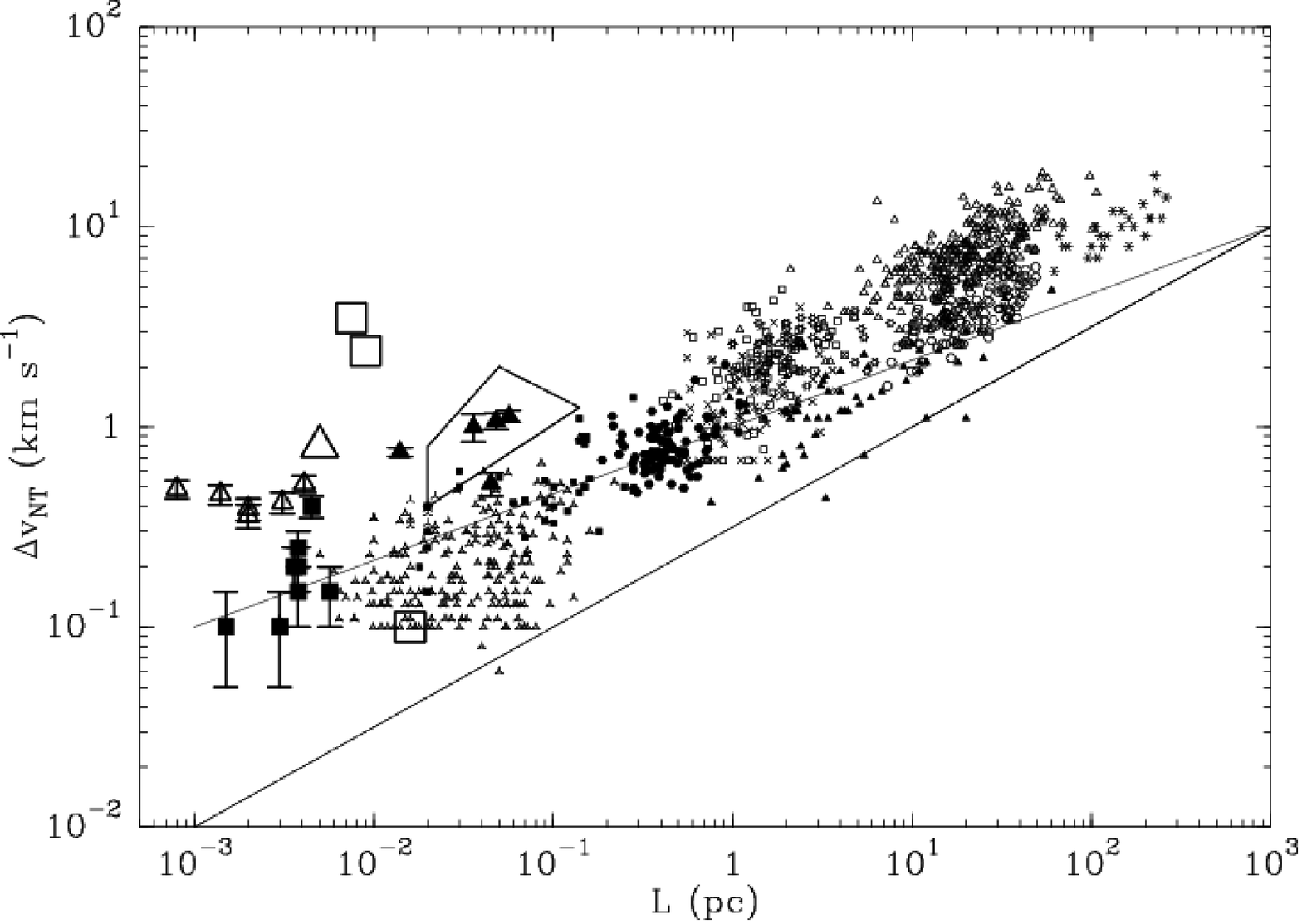}
  \caption{Size-linewidth relation for a large sample of \twCO(1-0) structures
    (see Appendix B) to which are added: the SAMS data (single-dish
    data from Heithausen (2002, 2006) (solid triangles), PdBI data
    within SAMS2 \citep{heithausen04} (open triangles)), a polygon that
    provides the range of values for the 12 structures of 
    \cite{sakamoto03} and the eight structures of Table 2
    (solid squares). The 3 empty squares without error bars 
show where the 
three pairs of PdBI-structures would be 
if not resolved spatially (\ie\ the velocity increment between the two structures 
would appear as a linewidth for the pair). Same with the large triangle
for the pair of structures in SAMS2. 
    The straight lines show the slopes 1/3 and 1/2 for comparison.  }
  \label{fig:vntr}
\end{figure}

Our results broaden the perspective regarding the existence of
small-scale CO structures in molecular clouds.  Heithausen (2002,
2004, 2006) has found small-area molecular structures (SAMS) 
that are truly isolated CO features in the high latitude sky. 
PdBI observations of the SAMS 
(Heithausen 2004) reveal bright sub-structures that are all brighter
and broader than our PdBI-structures. Unfortunately, the emission has 
been decomposed into clumps, a questionable procedure 
because short-spacings have not been combined to the PdBI
data and the CLEAN procedure tends to create structures on the
beam scale. The large \HH\  densities inferred are therefore likely overestimated.
An interesting feature can be seen in the channel maps, though.
Two elongated
thin patterns cross the field, reminiscent for their thickness
and length of what is found in the present study. A velocity shear of
180 \kmspc\ is estimated between these two elongated structures for a
velocity separation of 0.9 \kms\ and a {\it pos} spatial separation of $10~''$
on average (or 5~mpc at the assumed distance of 100~pc).  
This velocity-shear is thus commensurable with the
two largest values found in the Polaris field.

Ingalls et al. (2007) have detected milliparsec clumps in a high
latitude cloud. They are located in the line-wings of the CO
single-dish spectrum and they model them as tiny (1-5 mpc) clumps of density 
of a few 10$^3$ \cc. A more detailed comparison with the present results is not possible because  
they do not analyze individual structures.

Sakamoto \& Sunada (2003) have discovered a number of CO
small-scale structures in the low-obscuration regions of 
long strip maps beyond the edge of the
Taurus molecular cloud. Their main characteristics are their large
line-width and their sudden appearance, and disappearance, within
0.03 to 0.1 pc. The authors interpret these features as the
signature of structure formation induced by the thermal instability of
the warm neutral medium (WNM) in the turbulent cloud envelope. 
These CO small-scale
structures thus carry the kinematic signatures of the embedding WNM, hence
their large velocity dispersion, both interclump and
intraclump. The inferred line ratio, $R(2-1)/(1-0)=0.4$, is low, 
consistent with a low excitation temperature and \HH\
densities lower than $\sim 10^2$\cc.  The authors  propose that their
small-scale CO structures pinpoint molecule-forming regions, driven by
the thermal instability in the turbulent diffuse ISM.  

Our data therefore share many properties with these different samples.
Fig.\ref{fig:vntr} allows a comparison of the projected size and
linewidth of the above milliparsec-scale structures with those of
\twCO(1-0) structures identified in data cubes from non-star-forming
regions of all sizes, up to several 100~pc (see the relevant
references in Appendix B).  Although some of them (a few individual
PdBI-structures of our sample) further extend the general scaling law
down to 2~mpc, most of them significantly depart from this law by a
large factor.  As already mentioned in Section 5.1, the departure is
the largest for the pairs of PdBI-structures, as they would appear if
they were not resolved spatially \ie\ as anomalously broad structures
with respect to their size.  The increased scatter of velocity-widths
of the structures below 0.1~pc down to 1~mpc may be seen as another
manifestation of the intermittency of turbulence in translucent
molecular gas.

\section{Comparison with experiments, numerical simulations and chemical models}

 The present data set 
   discloses small-scale structures of intense velocity-shears that carry the 
statistical properties of intermittency and, in conjunction with that of HF09, reveals
 a connexion between parsec-scale and  milliparsec
   scale velocity-shears. The dynamic range of coupled scales in the Polaris Flare is
   therefore on the order of $\sim 10^3$. Moreover, velocity
   differences, up to 3.5~\kms, close to the rms velocity dispersion of the CNM
   turbulence measured on 10-pc scales (or
   more) \citep{mamd03,haud07}, are found in the PdBI field over $\sim$ 10 mpc, without
   any detected density enhancement nor shock signature.  We argue
   that the SEE(D)S are the CO-rich parts of straining sheets
in a gas undetected in \twCO(1-0), likely the CNM,
 and that the fast CO enrichment is driven by enhanced turbulent
 dissipation in the intense velocity-shears. We show
 below that these findings may be understood in the light of recent
 numerical simulations of incompressible and compressible turbulence,
 and the TDR chemical model of \cite{godard09}.

 The fact that the most dissipative structures appear to
  be layers of intense strain-rate is consistent with recent results of numerical
  simulations of incompressible turbulence at high Reynolds number
  \citep{moisy04} and laboratory experiments \citep{gana08}.
These regions are not randomly distributed and form 
inertial-range clusters \citep{moisy04} or develop 
at the boundaries regions of high level of vorticity 
(\ie\ vortex tubes) \citep{gana08}.
Coupling between small-scale statistics of the velocity field and 
the properties of the large-scale flows is also
  clearly probed in the high-$Re$ 
  numerical simulations of \cite{mininni06a}: correlations are observed between large-scale
shear and small-scale intermittency.

In compressible turbulence, the fact that the most dissipative structures
are shear-layers  is not expected.
Yet, in their hydrodynamical simulations of mildly compressible turbulence,
\cite{porter02} show that the compressible component of the velocity
field is weaker than its solenoidal counterpart by a factor $\sim 3$,
independent of the nature of the driving process (compressible or
solenoidal) and \cite{vestuto03} find that the energy fraction in the
solenoidal modes is dominant and increases with the magnetic field
intensity in compressible magneto-hydrodynamical (MHD) turbulence. 
These numerical experiments
are still far from approaching the ISM conditions but they suggest
that turbulent dissipation may occur primarily in solenoidal modes,
\ie\ without direct gas compression, and that the properties of the small scales are 
coupled to the large-scales.

 In the TDR models of \cite{godard09}, the chemical enrichment of
  the CNM is driven by high gas temperatures and enhanced ion-neutral
  drift, without density enhancement.  The temperature increase is due
  to viscous dissipation in the layers of largest velocity-shears at
  the boundaries of coherent vortices\footnote{ the ``sinews of turbulence''
    put forward by \cite{moffatt94} that link large-scale strain and
    small-scale vorticity}. The large ion-neutral drift occurs in the
  layers of largest rotational velocity in which ions and magnetic
  fields decouple from neutrals.  These two dissipative processes
  trigger endothermic chemical reactions, blocked at the low
  temperature of the CNM. Enrichments consistent with observations are
  obtained for turbulent rates-of-strain $a= 10^{-11}$ s$^{-1}$
  induced by large scale turbulence and for moderately dense gas
  ($n_{\rm H} < 200$ \cc) characteristic of the CNM.  There is no
direct determination of the rates-of-strain generated by turbulence in
the CNM. We note however that the largest observed velocity-shear
(Table 3) corresponds, if the projected quantities provide
reasonable estimates, to a comparable rate-of-strain.  The range of
observed CO column densities from $N({\rm CO}) = 10^{14}$ to $1.6
\times 10^{15}$\cq\ can be reproduced by intense velocity-shears
occurring in gas of density 100 to 200 \cc.  In this framework, the
energy source tapped to  enrich the medium in molecules is the
supersonic turbulence of the CNM.
  
The association between the large observed velocity-shears and local
enhanced dissipation rate is therefore supported not only by the earlier works
presented in the Introduction but also by a quantitative agreement
between the TDR chemical models and the present
observational results.  We cannot rule out however a contribution of
low-velocity C-shocks to the turbulent dissipation.  If they propagate
in the CNM, they are not visible in the CO lines. Such shocks are not
yet reliably modelled (Hily-Blant et al. in preparation).

\section{Conclusions and perspectives}

IRAM-PdBI observations of a mosaic of 13 fields in the turbulent
environment of a low-mass dense core have disclosed small and
weak \twCO(1-0) structures in translucent molecular gas.
They are straight and elongated structures 
but they are not filaments because, once merged with short-spacings data, the PdBI- 
structures appear as the sharp edges of larger-scale structures. 
Their thickness is as small as $\approx$ 3 mpc (600 AU), and their length, up to 70
mpc, is only limited by the size of the mosaic. 
Their CO column density is a well determined
quantity for the excitation conditions found at larger scale and is in
the range $N({\rm CO)}=10^{14}$ to $10^{15}$\cq.  Their \HH\ density,
estimated in several ways, including the continuum emission of the
brightest structure, does not exceed a few 10$^3$~\cc.
Their well-distributed orientations
can be followed in the larger-scale environnement of the field. Six of them form three pairs
of quasi-parallel structures, physically related. The velocity-shears estimated for the
three pairs include the largest ever measured in non-star-forming clouds (up to 780 \kmspc).  

The PdBI-structures are therefore not isolated and are the edges of
so-called SEE(D)S for sharp-edged extended (double) structures.  We
show that the SEE(D)S are thin layers of CO-rich gas and that their
sharp edges pinpoint a  small-scale dynamical process, at the origin of the CO
contrast detected by the PdBI.  We propose
 that the SEE(D)S are the outcomes of the chemical enrichment
driven by intense  dissipation occurring in large  
velocity-shears and that they are CO-rich layers 
swept along by the straining field of CNM turbulence.

The present work is the first detection of mpc-scale intense velocity-shears
belonging to a parsec-scale shear. 
The large departure from average of the kinematic properties of these structures, 
confirms that they are a manifestation of the small-scale intermittency of turbulence 
in this high latitude field, a property  already established on statistical grounds (HF09). 
The values of the velocity-shears (or rate-of-strain)
provide a quantitative constraint on the dissipation rate 
 that can be  compared to chemical models. 
The link between the turbulent dissipation in the diffuse gas and the dense core
observed in the vicinity of the PdBI mosaic (Fig.~\ref{fig:location}) still remains to be
established.

Last, we would like to stress that sub-structure still exists 
in these mpc-scale structures of the diffuse
  ISM and that the next generation of interferometers (e.g. ALMA)
should be able to observe gas  at the dissipation scale of
  turbulence (that is still unknown) or at least observe the effects
  on the ISM (temperature, excitation, molecular abundances) of the
  huge release of energy expected to occur there.

\begin{acknowledgements}
  We thank the IRAM staff at Plateau de Bure and Grenoble for their
  support during the observations. EF is most grateful to Michael
  Dumke, Emmanuel Dartois, Anne Dutrey and St\'ephane Guilloteau for
  their help during the early stages of the data reduction. EF also acknowledges
the stimulant discussions over the years with E. Ostriker, P. Hennebelle, A. Lazarian, 
B.G. Elmegreen, M.M. Mac-Low, E. Vasquez-Semadeni and many others that cannot be 
listed here.   
 We thank J. Scalo, our (formerly anonymous) referee, for his 
dedicated efforts at making us write our observational paper accessible to numericists. 
\end{acknowledgements}

\bibliographystyle{aa}
\bibliography{10963}

\appendix
\section{Noise level in the mosaic}

\begin{figure}
  \centering
 \includegraphics[width=0.8\hsize{},angle=-90]{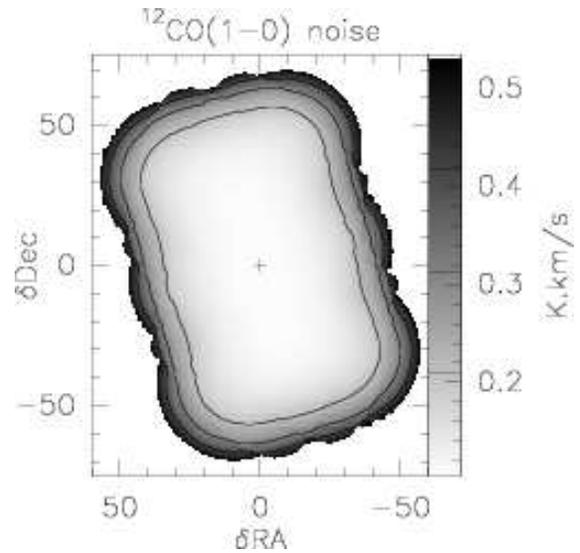}
  \caption{Map of the noise level in \Kkms\ over the 13-field mosaic. }
  \label{fig:noise}
\end{figure}

  Mosaic noise is inhomogeneous due to primary beam correction. This 
 is shown in Fig.\ref{fig:noise}. The 13-field mosaic 
produces a large area with
uniform noise level. Only at the edge of the mosaic  does it 
increase sharply due to the 
primary beam correction (the contour shown are at a 2, 3 and 4 sigma
level, 1 sigma being measured at the map center on a channel devoided
of signal).

\section{The size-linewidth scaling law}

Molecular cloud parameters have long been determined 
as those of three--dimensional structures
isolated in the four--dimensional space of the molecular line data
sets $T_L(x,y,v_z)$, the line brightness temperature being a function 
of position in the {\it pos} (two coordinates $x,y$), and one
spectral dimension, the projected velocity on the {\it los}
direction $v_z$. In this 4D space, 3D structures are isolated following
different methods \citep{stutgust90,williams94,falgarone87,loren89a,
falgarone92}.
The size and linewidth of the large number of clouds displayed in 
 Fig.\ref{fig:vntr} have been obtained 
by using published values, corrected in several cases to 
allow the size and linewidth obey the same definitions 
in all the samples \citep[see][]{falgahouch98}.
The structures are identified in \twCO(1-0) molecular line surveys of
the central parts of the Galaxy (stars, 
\cite{dame86}, open triangles \cite{solomon87}) and of the third quadrant
(open hexagons, \cite{may97}), in the Rosette (crosses) and Maddalena
(open squares) molecular clouds \citep{williams94}, 
in non--star-forming clouds 
(solid triangles, \cite{falgarone87}, solid squares, \cite{falgarone92}, 
tripods,
\cite{lemme95}), in $\rho$ Ophiuchus (solid hexagons, \cite{loren89a}) and in a
high latitude cloud (starred triangles, \cite{heithausen98}).

\end{document}